\documentclass[11pt,a4paper,fleqn]{article}
\usepackage[T1]{fontenc}
\usepackage{lmodern}
\AtBeginDocument{\DeclareFontShape{T1}{lmr}{m}{scit}{<->ssub*lmr/m/scsl}{}}
\usepackage[margin=25mm]{geometry}
\usepackage[numbers,square]{natbib}
\usepackage{amsmath,amsthm,amsfonts,amssymb,mathtools}
\usepackage{graphicx}
\usepackage{enumerate}
\usepackage{tikz}
\usetikzlibrary{positioning,calc,shapes,decorations.markings}
\usepackage{float,caption,subcaption,placeins}
\usepackage{algorithm,algpseudocode}
\usepackage{xspace}
\usepackage{microtype}
\usepackage[hidelinks]{hyperref}
\hypersetup{
  pdftitle={Balanced Divide and Conquer for Rooted Maximum Agreement Forests},
  pdfauthor={Feng Shi, Yexin Peng, Xin Zeng, Jingyi Liu, Qilong Feng, Jianxin Wang}
}
\providecommand{\algname}[1]{\textnormal{\textsc{#1}}}
\newcommand{\af}{\textsc{AF}\xspace}
\newcommand{\afs}{\textsc{AFs}\xspace}
\newcommand{\maf}{\textsc{MAF}\xspace}
\newcommand{\mafs}{\textsc{MAFs}\xspace}
\newcommand{\FAIL}{\textnormal{\textsc{Fail}}}
\DeclareMathOperator{\ord}{ord}
\DeclareMathOperator{\op}{opt}
\DeclareMathOperator{\lca}{lca}
\newtheorem{theorem}{Theorem}
\newtheorem{lemma}[theorem]{Lemma}
\newtheorem{definition}[theorem]{Definition}
\newtheorem{corollary}[theorem]{Corollary}

\title{Balanced Divide and Conquer for Rooted Maximum Agreement Forests}
\author{%
Feng Shi\thanks{Corresponding author: \href{mailto:fengshi@csu.edu.cn}{\texttt{fengshi@csu.edu.cn}}.
ORCID: \url{https://orcid.org/0000-0002-1415-0515}.},
Yexin Peng, Xin Zeng, Jingyi Liu,\\
Qilong Feng, Jianxin Wang\\[0.6em]
\small School of Computer Science and Engineering,\\
\small Central South University, Changsha 410083, China}
\date{}

\begin{document}
\maketitle

\begin{abstract}
We develop a balanced divide-and-conquer framework for the maximum
agreement forest problem on rooted binary phylogenetic tree--forest
instances. The goal is a common subforest with the fewest components;
budget $k$ bounds the increase relative to the input forest.
We balance edges associated with an unknown optimum to obtain three
subproblems with smaller budgets. Branching separates the inputs,
while temporary marker leaves allow a global optimum to be recovered
from independently computed optimal solutions.
For any fixed $B>\varphi=(1+\sqrt5)/2$, the framework converts a
tree--forest solver running in $O^*(B^k)$ time into one running in
$O^*((\sqrt{\varphi B})^k)$ time. Starting from an
$O^*(2.3391^k)$-time solver, two applications yield an
$O^*(1.7743^{k^*})$-time fixed-parameter algorithm for two trees,
where $k^*$ is the number of components in a maximum agreement forest 
minus one. For every fixed $\varepsilon>0$,
a fixed number of applications yields $O^*((\varphi+\varepsilon)^{k^*})$
time. Here $O^*$ suppresses polynomial factors, whose degree and
constants may depend on $\varepsilon$.
\end{abstract}


\section{Introduction}
\label{sec:intro}

Phylogenetic trees are widely used to represent evolutionary relationships among a set of extant species, or more generally taxa~\cite{huelsenbeck2002inferring,kapli2020phylogenetic,woese2000interpreting}.  Each leaf represents an observed taxon, whereas internal vertices represent hypothetical common ancestors or evolutionary divergence events.  A rooted phylogenetic tree additionally specifies an ancestral direction: the root represents a common ancestor and the paths from the root toward the leaves describe the direction of evolutionary descent.  By contrast, an unrooted phylogenetic tree records relationships among the taxa without specifying which vertex should be interpreted as the common ancestor.

Different parts of a genome need not support the same phylogenetic tree.  In particular, evolutionary processes such as horizontal gene transfer, hybridization, and recombination may lead to different tree topologies for different homologous genomic regions~\cite{gallardo2017phylogenetics,gontier2015reticulate,nakhleh2004reconstructing}.  Comparing phylogenetic trees and quantifying their structural disagreement is therefore an important problem in computational phylogenetics.  A number of distance measures have been proposed for this purpose, including the Robinson--Foulds distance~\cite{1}, nearest-neighbor interchange (NNI) distance~\cite{2}, tree bisection and reconnection (TBR) distance, subtree prune and regraft (SPR) distance~\cite{3,4}, hybridization number~\cite{hybridization}, and maximum parsimony distance~\cite{jones2021maximum}.

To compare two phylogenetic trees, we assume that their leaves represent the same set of taxa.  Let $X$ denote this common set of taxa (equivalently, leaf labels), where $|X| = n$, and let $T_1$ and $T_2$ be two rooted binary phylogenetic trees whose leaves are labeled by $X$.  Informally, an \emph{agreement forest} (\af) describes evolutionary structure that can be found simultaneously in both trees after a number of edges have been removed.  More precisely, its components partition the label set $X$, and for every component label set, the corresponding restricted subtree has the same topology in $T_1$ and $T_2$; moreover, the subtrees representing different components are vertex-disjoint in each input tree.  Equivalently, one may view an \af as a common forest obtained from the two input trees by deleting edges and applying the usual reductions to the resulting unlabeled vertices.  The \emph{order} of a forest is the number of its components, and we denote it by $\ord(F)$ for a forest $F$.  A \emph{maximum agreement forest} (\maf) is an \af of minimum order~\cite{7}; we denote this minimum order for $(T_1,T_2)$ by $m(T_1,T_2)$. Thus, despite the terminology ``maximum'', the optimization objective is to minimize the number of components, or equivalently to preserve as much common structure of the two input trees as possible.  Throughout the paper, \maf denotes a maximum agreement forest as a solution object; the corresponding computational problem is referred to as the \maf problem.

Among the distance measures mentioned above, TBR and rooted SPR (rSPR) distances are particularly closely related to \afs.  For two unrooted binary phylogenetic trees, the analogous unrooted notion of an \maf characterizes TBR distance: if an \maf has order $m$, then the TBR distance is exactly $m-1$~\cite{6}.  For two rooted binary phylogenetic trees, the corresponding relationship holds for rSPR distance under the standard convention in which a new vertex labeled $\rho$ is attached above the original root of each tree and becomes the new root.  If an \maf of the resulting $\rho$-augmented pair has order $m$, then the rSPR distance of that pair is $m-1$~\cite{7}.  Hence an \maf provides a combinatorial route to computing these edit distances, and algorithmic improvements for the \maf problem translate directly into improvements for the corresponding distance computation under the appropriate convention.

The \maf problem is NP-hard for two rooted binary phylogenetic
trees~\cite{7}; for two unrooted binary phylogenetic trees, it is both
NP-hard and MAX SNP-hard~\cite{5}. This has motivated extensive research on approximation algorithms, fixed-parameter algorithms, kernelization, and practical decomposition techniques.
In this paper we focus on the \maf problem for two rooted binary phylogenetic trees.  Several related variants have also been studied, including the \maf problem for two unrooted binary trees~\cite{6,Hallett,mestel2026split,whidden2018calculating,12}, for two rooted multifurcating trees~\cite{8,17}, for two unrooted multifurcating trees~\cite{14,Yang}, for multiple binary trees~\cite{15,20,18,19}, and for multiple rooted multifurcating trees~\cite{shi2018jcss}.

\subsection{Related Work}

The literature considers rooted phylogenetic trees both with
and without a distinguished root label $\rho$.
The main design ideas behind the algorithms reviewed below
are essentially independent of this choice.
We therefore discuss these results together and specify
the convention used in each paper where it matters for
the comparison.

{\bf Approximation Algorithms}.
The approximation guarantees below are stated for the cut-count
objective, defined as the \af order minus one.
The first correct constant-factor approximation algorithm for the
\maf problem was given by Bonet et al.~\cite{9}, who
obtained an approximation ratio of $5$.\footnote{Earlier
claimed approximation bounds for related agreement-forest
problems~\cite{5,8} had gaps in their analyses; see the discussion
and counterexamples in~\cite{9}.}
Bordewich et al.~\cite{10} improved the ratio to $3$ with
running time $O(n^5)$.  This approximation ratio was retained while
the running time was subsequently improved: an $O(n^2)$
3-approximation was given in~\cite{11}, and Whidden and
Zeh~\cite{12} obtained a linear-time 3-approximation.
Further progress focused on improving the approximation ratio.  Shi
et al.~\cite{shiapprox} obtained a $2.5$-approximation in
$O(n^2)$ time, and Chen et al.~\cite{chen2016improved} improved
the ratio to $7/3$, again with quadratic running time.  Building on
the structural analysis underlying the $7/3$-approximation, Chen
et al.~\cite{chen2017new} obtained a 2-approximation with cubic
running time.  Independently, Schalekamp et al.~\cite{Schalekamp}
derived a 2-approximation using LP duality.  Although the original
presentation did not emphasize the implementation complexity, the
full version~\cite{olver2022duality} showed that this approach can be
implemented in $O(n^2)$ time.

{\bf Fixed-Parameter Algorithms}.
The \maf problem has been extensively studied from the
viewpoint of parameterized complexity, with the order of an \maf, 
or equivalently, that order minus one, as the parameter.
Throughout this overview, we use $k$ to denote the minimum order
minus one.
The goal is to obtain algorithms with running time
$f(k)\cdot n^{O(1)}$, where $f$ is a computable function depending
only on $k$~\cite{fptbook}.
We use $O^*(\cdot)$ to suppress factors polynomial in the input size.
Most fixed-parameter algorithms for the \maf problem follow
a branching paradigm.
When the two input structures disagree locally, the algorithm
identifies a small collection of edge-deletion alternatives such
that at least one alternative is compatible with an \maf.
It branches on these alternatives, decreases the remaining cut
budget, and recursively solves the resulting instances.
The exponential base is governed by the resulting branching
recurrences, which depend on both the number of alternatives and
the budget consumed by each alternative.

Bordewich et al.~\cite{10} gave an $O^*(4^k)$-time algorithm.
Whidden et al.~\cite{siam} improved the bound to
$O^*(2.42^k)$ through a refined analysis of local configurations,
including sibling pairs and pendant subtrees.
Chen et al.~\cite{zhizhongchen} further analyzed these
configurations and their interactions, obtaining an
$O^*(2.344^k)$ bound.
These developments illustrate a central direction in earlier work:
stronger branching rules and finer case analyses progressively
reduce the exponential base.
Whidden's PhD thesis~\cite{whidden2013efficient} also states an
$O^*(2^k)$ bound. Since we are not aware of a peer-reviewed
publication establishing this bound, we do not use it in our
subsequent analysis.

Mestel et al.~\cite{mestel2026split} developed a
split-or-decompose algorithm for the \maf problem.
On a tree--forest instance $(T,F)$ with additional-cut budget
$\kappa$, their algorithm seeks an \af of order at most
$\ord(F)+\kappa$ and runs in $O^*(2.3391^\kappa)$ time.
It maintains one input as a reference tree $T$ and represents
the other by a forest $F'$ obtained through edge deletions.
When components of $F'$ induce overlapping subtrees in $T$,
these overlaps prevent a direct decomposition into independent
subinstances. Splitting and branching rules are applied to
resolve the overlaps and the remaining rooted configurations.
Once the induced subtrees are pairwise vertex-disjoint, each
component defines a separate subinstance, and solutions to
these subinstances can be combined into a solution for the
current input.
The analysis also uses information about how the remaining
work is distributed among the subinstances.
A bounded-depth probing procedure tests each subinstance with
a constant cut budget, solving those within that budget and
obtaining lower bounds for the others.
The cuts required by other subinstances can then be deducted
from the budget supplied to a recursive call.
This gives a running-time gain when several subinstances
require nontrivial work, while additional branching rules
handle the limiting configurations.

The split-or-decompose framework is particularly relevant to our work
because it shows that decomposition can contribute directly to the
exponential analysis of the \maf problem, rather than serving
only as a practical preprocessing technique. In that framework,
however, decomposition becomes available only after branching has
eliminated overlaps among the current forest components.  Moreover,
the framework does not impose a balance condition on the remaining
budgets of the resulting subinstances.  Consequently, a decomposition
can in principle be highly unbalanced, with one subinstance retaining
almost all of the remaining budget.

{\bf Kernelization}.
A kernelization algorithm transforms an instance in polynomial time
into an equivalent instance whose size is bounded by a function of
the parameter.  Although kernelization does not improve the
exponential base of a fixed-parameter algorithm, it can bound the size of the
instance passed to the exponential-time phase by a function of the
parameter.  For the standard $\rho$-augmented formulation with rSPR
parameter $d$, Bordewich and Semple~\cite{7} obtained a kernel with at
most $28d$ taxa.  For $d\geq1$, Kelk {\it et
al.}~\cite{kelk2023cyclic} subsequently improved this bound to
$9d-3$ taxa using subtree, chain, and $3$-$2$-chain reductions; the
case $d=0$ is trivial.
In this paper, we use these reductions only as a preprocessing step on
the original tree--tree instance.  Their compatibility with the root
convention used in our technical development, as well as the resulting
size bound for the recursive instances, is established in
Section~\ref{sec:root-kernel-discussion}.

{\bf Cluster Partitioning}.
Linz and Semple~\cite{linz2011cluster} introduced cluster partitioning
as a practical decomposition technique for the rooted \maf problem.  A
suitable cluster shared by the two input trees can be used to reduce
the computation to smaller subproblems whose solutions are
subsequently combined according to the cluster-reduction framework.
This approach is effective in practice because phylogenetic instances
often contain substantial common structure, allowing the
exponential-time solver to operate on considerably smaller
subproblems~\cite{li2017computing,whidden5}.
Cluster partitioning, however, does not by itself guarantee an
improvement in the worst-case exponential base.  An input pair need
not contain a useful common cluster, and even when such a
decomposition exists, the parameter may be distributed arbitrarily
unevenly among the resulting subproblems.

\subsection{Our Contributions}

Motivated by the split-or-decompose strategy~\cite{mestel2026split}
and cluster partitioning~\cite{linz2011cluster}, we develop a
balanced-center framework for the \maf problem on two rooted binary
phylogenetic trees. The framework decomposes a tree--forest instance
into three side instances, solves them independently, and reconstructs
an \maf for the original instance. A subproblem on relatively few
labels may still require most of the budget for additional components.
We therefore seek a partition that balances the budgets needed to
solve the three subproblems. Finding and enforcing such a partition
presents two difficulties. The desired balance depends on an unknown
optimal solution, and a component of the input forest or of that
solution may contain labels from more than one part. The following
three ingredients address these difficulties and yield an improved
running-time bound.

To state the balance guarantee, we use a refined parameterization
for forest pairs. For $(F_1,F_2)$, let $m(F_1,F_2)$ denote the minimum
order of an \af. A set of labels is \emph{separating} if every two
distinct labels belong to different components in at least one input.
Let $\sigma_{\max}(F_1,F_2)$ be the maximum size of such a
set~\cite{shi2018jcss}; this is a lower bound on the order of every
\af. Given a budget $k$, the task is to determine whether
$m(F_1,F_2)\le \sigma_{\max}(F_1,F_2)+k$.
For a fixed target \af $G$, define its \emph{gap} by
$r=\ord(G)-\sigma_{\max}(F_1,F_2)$.
Thus $r\le k$ whenever $G$ is feasible for the supplied budget.
The algorithm uses $k$; the unknown target and its gap $r$ guide only
the analysis. On a tree--forest pair $(T,F)$, we have
$\sigma_{\max}(T,F)=\ord(F)$, so the gap is exactly the number
of additional cuts needed to obtain the target from $F$.

\begin{enumerate}
\item {\bf Finding a center that balances the remaining disagreement.}
Our first task is to identify where the unresolved disagreement lies.
Fix a tree--forest instance $(T,F)$ and a target \af $G$ with gap
$r$. We prove that this gap can be represented by a \emph{critical
edge-deletion set} of exactly $r$ edges in $T$: deleting these edges
and applying the greedy reduction process recovers $G$. The edges
also admit an ordering in which each prefix of $j$ deletions increases
$\sigma_{\max}$ by exactly $j$. This connects the numerical gap
to specific edge locations and allows us to bound the gap on each
side of a partition.

For $r\ge3$, a weighted-centroid argument gives a nonroot internal
vertex $c$ whose removal from $T$ gives three sides, each containing at most
$\lfloor r/2\rfloor$ critical edges. Consequently, each restricted
side target has gap at most $\lfloor r/2\rfloor$. Although the
algorithm does not know $G$ or its critical edges, it enumerates all
$O(n)$ possible centers and therefore includes a suitable choice.
Thus we obtain a partition that balances the remaining disagreement,
rather than merely the number of labels. Instances with gap at most
two are handled by the auxiliary solver.

\item {\bf Separating the inputs without losing an optimal solution.}
A balanced center $c$ partitions $T$, but the same label partition
need not separate $F$.
Let $X_0$ be the labels on the root side of $c$, and let
$X_1,X_2$ be the labels on the other two sides.
We enforce this partition through two binary separations:
first $X_1\mid(X_0\cup X_2)$, and then $X_2\mid X_0$, with these expressions referring to the inherited labels.
At each stage, we can separate the inputs without branching if
every component of the current forest has all its labels on one
side of the split, either already or after deleting a single edge.
Otherwise, we use a two-color branching procedure.

A further difficulty is that a component of the target \af 
may contain labels from both sides of a separation.
Keeping the sides permanently disconnected would split this
component and could therefore exclude the target optimum.
To preserve the possibility of recovering that solution, we
introduce a pair of marker leaves to record each temporarily cut
connector.
After the sides have been solved, the components containing these
markers determine whether the connector is restored.
The two separation stages are reconstructed in reverse order.

Preserving the option to reconnect is not sufficient by itself:
the reconstructed forest must also have the required order.
A target that uses a connector gains one component when that
connector is cut. A target that omits it gains two singleton
marker components instead. Reconstruction must compensate for
both possibilities, even though a side solver may return an
optimum different from the target restriction. We address this
by trying the constant number of choices of singleton markers.
For each choice we prepare ordinary tree--forest instances,
so the solver needs no additional marker constraints. For a
separation that preserves the fixed target \af $G$, the
singleton-marker choice prescribed by the transformed target
allows the returned side forests to be reconstructed into an
\af of order at most $\ord(G)$. Taking $G$ to be an \maf for
the original instance establishes global optimality.

\item {\bf Turning balanced decomposition into faster algorithms.}
The balance guarantee must hold for the marked instances actually
sent to the side solver. After singleton preparation, the child-side
target gaps retain their original bounds, while the root-side gap
can increase by at most one. Thus, on a
target-compatible branch and for the target-induced singleton pattern,
every prepared side target has gap at most
$\lfloor r/2\rfloor+1$. The remaining separation budget $b$ gives
an independent bound, so the algorithm uses
$\kappa=\min\{\lfloor k/2\rfloor+1,b\}$
as the supplied budget for each side call. Since every side is a
tree--forest instance, $\kappa$ is exactly the additional-cut budget
used by standard solvers.

The key to the running-time analysis is to account for separation
and side solving together: a branch that spends more deletions
enforcing the partition leaves less work for the side solver.
Let $\varphi=(1+\sqrt5)/2$. We prove that a compatible side
solver with running time $O^*(B^\kappa)$, where $B>\varphi$, yields
an overall algorithm with exponential base $\sqrt{\varphi B}$.
The extra unit in the side budget bound $\lfloor k/2\rfloor+1$
covers the possible increase in the root-side gap due to the
markers. Marker operations also add at most two units to the
separation budget. If the framework is applied a fixed number of
times, these extra units affect only the constant factor in the
running-time bound.

By Theorem~\ref{thm:BC-Maf}, applying the framework with the
$O^*(2.3391^\kappa)$ solver of Mestel et al.~\cite{mestel2026split}
gives an $O^*(1.9455^k)$-time algorithm.
This algorithm satisfies the same input and output guarantees
as the original side solver. Applying the framework again with
the new algorithm as the side solver gives $O^*(1.7743^k)$ time.
For every fixed $\varepsilon>0$, repeating this process a suitable
fixed number of times gives an $O^*((\varphi+\varepsilon)^k)$-time
algorithm. The polynomial degree and hidden constant may depend
on $\varepsilon$.

\end{enumerate}

For each fixed number $x$ of applications of the framework, the
resulting algorithm runs in $O^*(\beta_x^k)$ time, where
$\beta_x=\varphi^{1-2^{-x}} \cdot 2.3391^{2^{-x}}$.
The degree of the polynomial factor suppressed by $O^*$ and the
hidden constant may both depend on $x$.
We apply the kernelization only to the original
tree--tree instance to control the polynomial
dependence on the input size.
On that instance, $\sigma_{\max}(T_1,T_2)=1$, so the optimum gap is
$k^*=m(T_1,T_2)-1$. The following theorem states the resulting
optimization bound in terms of $k^*$, as distinct from the supplied
budget $k$ used by the algorithm.

\begin{theorem}
\label{thm:final-2}
Let $T_1$ and $T_2$ be two rooted binary phylogenetic trees on the
same set of $n$ labels, and let $k^*=m(T_1,T_2)-1$.
An \maf for $(T_1,T_2)$ can be computed in
$O^*(1.7743^{k^*})$ time.  With the one-time kernelization used in
this paper, the running time is
$O(1.7743^{k^*} \cdot (k^*+1)^{O(1)}+n^3)$.
Moreover, for every fixed $\varepsilon>0$, there is a
fixed integer $x=x(\varepsilon)$ such that applying the
balanced-center framework $x$ times gives an algorithm
with running time $O^*((\varphi+\varepsilon)^{k^*})$,
where $\varphi=(1+\sqrt5)/2$.
The polynomial degree and hidden constant may depend
on $\varepsilon$.
\end{theorem}

To apply our algorithm to the standard rSPR formulation, we
use Lemma~\ref{lem:root-leaf-encoding} to encode the distinguished
root label $\rho$ as a leaf label.
The encoded pair has the same \maf order
as the pair with the labeled root $\rho$.
Hence its \maf order minus one equals
the rSPR distance of that rooted pair.
Applying our algorithm to this encoding gives the following
corollary.

\begin{corollary}
\label{cor:rspr-main}
Let $T_1$ and $T_2$ be two rooted binary phylogenetic trees, and for
$i\in\{1,2\}$ let $T_i^\rho$ be obtained by attaching a new vertex
labeled $\rho$ above the original root of $T_i$ and taking it as the
new root. Let $d=d_{\rm rSPR}(T_1^\rho,T_2^\rho)$.
Then an \maf for $(T_1^\rho,T_2^\rho)$ can be computed in
$O^*(1.7743^d)$ time.  More generally, for every fixed
$\varepsilon>0$, it can be computed in
$O^*((\varphi+\varepsilon)^d)$ time, where $\varphi=(1+\sqrt5)/2$.
\end{corollary}

{\it Organization.}
Section~\ref{sec:pre} introduces the definitions, parameters, and
requirements on the side solver.
Section~\ref{sec:Correctness} develops the critical-edge
representation, which is used in Section~\ref{sec:balanced-centers}
to establish the balanced-center property and bound the gaps of
the restrictions to the three label sets.
Sections~\ref{sec:easy} and~\ref{sec:difficult} describe the
separation procedures and show how the resulting forests can be
reconstructed.
Section~\ref{sec:balanced-center-algorithm} assembles these
procedures into the full algorithm, bounds the effect of fresh
markers on the side parameters, and proves correctness and
running time, including recursive applications of the algorithm.
Section~\ref{sec:root-kernel-discussion} discusses kernelization.

\section{Preliminaries and Problem Formulations}
\label{sec:pre}

We follow the conventions for rooted trees and forests used by
Mestel et al.~\cite{mestel2026split}. Let $X$ be a nonempty finite set of
labels.  For $|X|\geq2$, a \emph{rooted binary phylogenetic tree} on
$X$, or an \emph{$X$-tree}, is a directed tree whose leaves are
bijectively labeled by $X$.  Its root has in-degree~$0$ and
out-degree~$2$, every other internal vertex has in-degree~$1$ and
out-degree~$2$, and every leaf has in-degree~$1$ and out-degree~$0$.
For $|X|=1$, the unique $X$-tree consists of the single labeled
vertex.  All edges are directed away from the root and hence induce an
ancestor--descendant relation. We identify a label with its leaf.
Two vertices are \emph{siblings} if they have the same parent.
A pair of sibling leaves is called a \emph{cherry}.

An \emph{$X$-forest} is a collection of rooted binary phylogenetic
trees whose label sets form a partition of $X$.  Equivalently, an
$X$-forest can be obtained from an $X$-tree by deleting edges and
applying \emph{forced contraction}. Forced contraction consists 
of the following two operations~\cite{10,Hallett}:
\begin{enumerate}[(1)]
  \item delete an unlabeled isolated vertex or an unlabeled vertex of
        undirected degree~1; and
  \item suppress an unlabeled vertex of in-degree~$1$ and
        out-degree~$1$, replacing its two incident edges by one
        directed edge.
\end{enumerate}
These operations are applied repeatedly until neither applies.
Every remaining component contains at least one label.
A singleton component consists of a single labeled vertex;
in every other component, the root has in-degree~$0$ and
out-degree~$2$.
A forest is \emph{irreducible} if neither operation applies.
We regard forests that differ only by forced contraction
as the same forest.
The \emph{order} $\ord(F)$ of a forest $F$ is its number
of components.

For a vertex $v$ of a forest $H$, the subtree rooted at
$v$ consists of $v$, all its descendants, and the edges
joining these vertices.
We call a subtree of this form a \emph{pendant subtree}.
We allow the whole component when $v$ is its root.
If $v$ is not the root, the pendant subtree is attached
to the rest of its component by the edge entering $v$.

For a subgraph $H$ of an $X$-forest, denote by $L(H)$ the labels
occurring in $H$. For a vertex $v$, denote by $L(v)$ its
descendant label set. For vertices $u,v$ in the same component of a forest $H$,
let $\lca_H(u,v)$ denote their least common ancestor. 
Unless stated otherwise, $n$ denotes the number of labels in the current input.
For a forest $H$ and a nonempty subset $Y\subseteq L(H)$,
the \emph{restriction} $H[Y]$ is obtained as follows.
In each component $C$ of $H$ with $L(C)\cap Y\neq\varnothing$,
retain the minimal connected subgraph spanning $L(C)\cap Y$.
Take the union of these subgraphs and apply forced contraction.
If $L(C)\cap Y$ contains a single label, its contribution
is just the corresponding labeled vertex.

Two $X$-forests are \emph{isomorphic} if there is a root- and
direction-preserving graph isomorphism between them that preserves all
labels.  An $X$-forest $F_1$ is an \emph{$X$-subforest} of another
$X$-forest $F_2$ if deleting edges from $F_2$ and applying
forced contraction yields a forest isomorphic to $F_1$.  
An \emph{agreement forest} (\af) for an
$X$-forest pair $(F_1,F_2)$ is a common $X$-subforest of $F_1$ and $F_2$.
A \emph{maximum agreement forest} (\maf) is an \af
of minimum order, and $m(F_1,F_2)$ denotes this minimum order.

Let $G$ be an \af for an $X$-forest pair $(F_1,F_2)$.
For each component $C$ of $G$ and each $i\in\{1,2\}$,
there is a unique minimal connected subgraph of $F_i$
spanning $L(C)$. If $C$ is a singleton, this subgraph
consists of its labeled leaf.
The union of these subgraphs, before forced contraction,
is the \emph{representation} of $G$ in $F_i$.
The same definition applies to any subforest of an input forest.
For each $i\in\{1,2\}$, an edge or vertex of $F_i$ is
\emph{contained in} $G$ if it belongs to the representation
of $G$ in $F_i$.
Let $E_{F_i}(G)\subseteq E(F_i)$ denote the edges of this representation.
An edge of $G$ may correspond to a path in $F_i$;
$E_{F_i}(G)$ records the edges of $F_i$ before forced
contraction.

For a fixed pair $(F_1,F_2)$ of $X$-forests, we sometimes
specify an \af by a partition $\mathcal P=\{Y_1,\ldots,Y_q\}$ of $X$.
Each block $Y_j$ is the label set of one component of
the \af.
For this partition to specify an \af, each $Y_j$ must
be contained in the label set of a single component
of each input, and the restrictions $F_1[Y_j]$ and
$F_2[Y_j]$ must be isomorphic rooted trees.
In addition, for each $i\in\{1,2\}$, the minimal connected
subgraphs of $F_i$ spanning $Y_1,\ldots,Y_q$ must be
pairwise vertex-disjoint.
Although the blocks have disjoint label sets, their
spanning subgraphs may share internal vertices, so this
additional condition is necessary.
When these conditions hold, the common rooted restrictions
on the blocks form an \af of order $q$.
In particular, a block $\{x\}$ represents the singleton
component consisting of the labeled vertex $x$.

For an edge set $E$ of a forest $F$, let $F \setminus E$
denote the forest obtained by deleting the edges of $E$ and
then applying forced contraction.
For a single edge $e$, we also use the notation $F-e$ for
deletion without forced contraction: all remaining vertices
and edges are retained until forced contraction is explicitly
requested.
An edge $e$ of an $X$-forest $F$ is \emph{essential} if
$\ord(F\setminus\{e\})=\ord(F)+1$.
An edge set $E$ is an \emph{essential edge set} (\emph{ee-set}) if
$\ord(F\setminus E)=\ord(F)+|E|$.
Every irreducible $X$-subforest $F'$ of $F$ can be written as
$F \setminus E$ for an ee-set $E$ of size $\ord(F')-\ord(F)$.
For a forest pair $(F_1,F_2)$, an ee-set $E$ of $F_i$ is
\emph{optimal} if $F_i\setminus E$ is an \maf for the pair.

\subsection{Root Convention and the rSPR Distance}
\label{sec:maf-rspr}

We can use the standard root augmentation to relate \afs to
rooted subtree prune and regraft (rSPR) distance.
Let $T_1,T_2$ be rooted $X$-trees, and choose a new label
$\rho\notin X$. For each $i\in\{1,2\}$, obtain $T_i^\rho$
by adding a new root labeled $\rho$ whose only child is
the original root of $T_i$.
The definitions of subforest, agreement forest, restriction,
and order extend to these trees by allowing the labeled root.
The classical characterization gives
$d_{\rm rSPR}(T_1^\rho,T_2^\rho) = m(T_1^\rho,T_2^\rho)-1$~\cite{7}.
The quantity $m(T_1,T_2)-1$, computed without this
augmentation, need not equal the rSPR distance.

Our framework uses the algorithm of Mestel et
al.~\cite{mestel2026split} as an auxiliary tree--forest
solver. We therefore adopt their conventions for rooted
trees and forests, with all labels placed at leaves.
To apply the solver to $(T_1^\rho,T_2^\rho)$,
Lemma~\ref{lem:root-leaf-encoding} encodes the labeled
root $\rho$ as a leaf and produces an ordinary
leaf-labeled pair $(\overline T_1,\overline T_2)$ satisfying
$m(\overline T_1,\overline T_2) = m(T_1^\rho,T_2^\rho)$.
Thus the \maf order of the encoded pair $(\overline T_1,\overline T_2)$, minus one, equals 
the rSPR distance between $T_1^\rho$ and $T_2^\rho$.
The encoding is applied once, before the algorithm begins.
During recursion, $\rho$ is treated as an ordinary leaf
label, and side instances are not augmented again.
After the algorithm returns a forest, the same lemma
converts it into an \af for $(T_1^\rho,T_2^\rho)$ in
polynomial time without changing its order.

\subsection{Problem Formulations}
\label{sec:problem-formulations}

The \maf problem asks for an \maf for an $X$-forest pair, where each input may have several components. The parameterized version of the problem can be formulated in several ways, depending on the lower bound from which the order of the desired \af is measured. The most direct formulation uses the order of the desired \af as the parameter.

\medskip
\noindent
\textsc{Parameterized MAF Version 1} (abbr.~\textsc{p-MAF$_1$}).
Given $(F_1,F_2)$ and an integer $k_1$, decide whether
$m(F_1,F_2)\leq k_1$ and, if so, return an \af of order
at most $k_1$.

\medskip

An alternative parameter measures the gap above a lower bound
on the \af order. Define $\ord(F_1,F_2)=\max\{\ord(F_1),\ord(F_2)\}$.
Since every \af for $(F_1,F_2)$ is a subforest of both inputs,
its order is at least $\ord(F_1,F_2)$.
We therefore measure the allowed order above this bound,
rather than using the total order as the parameter.

\medskip
\noindent
\textsc{Parameterized MAF Version 2}
(abbr.~\textsc{p-MAF$_2$}).
Given $(F_1,F_2)$ and an integer $k_2$, decide whether
$m(F_1,F_2)\leq\ord(F_1,F_2)+k_2$ and, if so, return an
\af of order at most $\ord(F_1,F_2)+k_2$.

\medskip

This formulation also has an edge-deletion interpretation.
Assume, without loss of generality, that
$\ord(F_1)\leq\ord(F_2)$.
Then \textsc{p-MAF$_2$} asks whether $F_2$ has an ee-set $E$
with $|E|\leq k_2$ such that $F_2\setminus E$ is an \af
for $(F_1,F_2)$. Indeed,
$\ord(F_2\setminus E)
  =\ord(F_2)+|E|
  =\ord(F_1,F_2)+|E|$.
Thus $k_2$ is a budget for additional cuts in the input
forest of larger order.
This viewpoint is used by existing branching algorithms
for the problem~\cite{zhizhongchen,mestel2026split,
whidden2013efficient,siam}.

However, the lower bound $\ord(F_1,F_2)$ is not always tight, as it does not capture separations
that are already forced jointly by the pair.  We account for these
separations through the following lower bound of Shi et
al.~\cite{shi2018jcss}.  A subset $S$ of the label set $X$ is a {\it separating set} for an $X$-forest pair $(F_1,F_2)$ if, for any two distinct labels $l_1,l_2\in S$, the labels $l_1$ and $l_2$ are in different components of at least one of $F_1$ and $F_2$. A {\it maximum separating set} for $(F_1,F_2)$ is a separating set of maximum size. The size of a maximum separating set for $(F_1,F_2)$ is denoted by $\sigma_{\max}(F_1,F_2)$.

As shown below, $\sigma_{\max}(F_1,F_2)$ is a computable lower bound on the order of any \af for $(F_1,F_2)$, and this lower bound is at least $\ord(F_1,F_2)$. Thus it gives a potentially sharper parameter for the problem.
For later use, define the \emph{component-pair partition}
\[
  \mathcal P(F_1,F_2)
  =\{L(A)\cap L(B): A\in F_1,\ B\in F_2, \ L(A)\cap L(B) \ne \varnothing\}.
\]
Here $A\in F_1$ and $B\in F_2$ range over components. Two labels
lie in the same block of this partition exactly when they belong to the same
component in each input.

\begin{lemma}
\label{maxss}
A maximum separating set of an $X$-forest pair $(F_1,F_2)$ can be
constructed in linear time. Moreover, $\sigma_{\max}(F_1,F_2)=|\mathcal P(F_1,F_2)|$.
\end{lemma}

\begin{proof}
A separating set contains at most one label from each
block of $\mathcal P(F_1,F_2)$, since labels in the same
block belong to the same component in both inputs.
Conversely, choosing one label from each block gives
a separating set: labels from distinct blocks belong
to different components in at least one input.
This proves the equality.
To construct such a set, traverse both forests to assign
an integer index to each component.
Group the labels by their pairs of component indices
using radix sorting, and choose one label from each group.
These steps take linear time.
\end{proof}

Choosing one label from each component of either input shows that
$\sigma_{\max}(F_1,F_2)\geq\ord(F_1,F_2)$.  The other inequality used
throughout the paper follows from the next result.

\begin{lemma}[\cite{shi2018jcss}]
\label{lem:j-new1}
If $S$ is a separating set for $(F_1,F_2)$ and $F$ is an \af
for the pair, then $\ord(F)\geq|S|$.
\end{lemma}

It follows that every \af $F$ satisfies
$\ord(F_1,F_2)\leq\sigma_{\max}(F_1,F_2)\leq\ord(F)$. Thus, using $\sigma_{\max}(F_1,F_2)$ as the lower bound gives a potentially smaller gap parameter. This leads to the following parameterized formulation.

\medskip
\noindent
\textsc{Parameterized MAF Version 3} (abbr.~\textsc{p-MAF$_3$}).
Given $(F_1,F_2)$ and an integer $k_3$, decide whether
$m(F_1,F_2)\leq \sigma_{\max}(F_1,F_2)+k_3$ and, if so, return an \af of order
at most $\sigma_{\max}(F_1,F_2)+k_3$.

\medskip

Let $F^*$ be an \maf for $(F_1,F_2)$.  The optimum gaps for the last
two formulations are
$\op_2(F_1,F_2)=\ord(F^*)-\ord(F_1,F_2)$ and
$\op_3(F_1,F_2)=\ord(F^*)-\sigma_{\max}(F_1,F_2)$.  Since
$\sigma_{\max}(F_1,F_2)\geq\ord(F_1,F_2)$, we have
$\op_3(F_1,F_2)\leq\op_2(F_1,F_2)$, and the inequality can be strict.
The refined gap discounts all separations already forced jointly by
the input forests and measures only the remaining unresolved work.

For a tree--forest pair $(T,F)$, we have
$\sigma_{\max}(T,F)=\ord(F)$, so
$\op_3(T,F)=m(T,F)-\ord(F)=\op_2(T,F)$. Hence the
additional-cut budget $\kappa$ supplied to a tree--forest solver is
also the refined budget for that call.  For a $\rho$-augmented
tree--tree pair, $\sigma_{\max}=1$, and therefore
$\op_3(T_1^\rho,T_2^\rho)=m(T_1^\rho,T_2^\rho)-1=d$, where
$d=d_{\rm rSPR}(T_1^\rho,T_2^\rho)$.
Thus the refined formulation agrees with the parameters used in both
settings while extending naturally to general forest--forest pairs.
In Section~\ref{sec:Correctness}, we represent a target's
refined gap by a critical ee-set. The locations of these
edges provide the basis for the balanced-center construction.

\subsection{Terminology and Parameter Convention}
\label{sec:terminology-basic}

Throughout the divide-and-conquer analysis, $k$ denotes
the supplied refined budget of the current instance,
and $\op_3$ denotes its optimum refined gap.
For a fixed target \af $G$ of the current pair $(F_1,F_2)$,
we write
$r=\ord(G)-\sigma_{\max}(F_1,F_2)$.
If $G$ is feasible for the supplied budget, then $r\leq k$.
During the two binary-separation stages, $b$ denotes
a temporary budget.
Each budget-charged edge deletion decreases $b$ by one,
while each new reconstruction record increases $b$ by one.
These updates are defined in
Section~\ref{subsec:separation-budgets}.
The symbol $\kappa$ denotes the additional-cut budget
passed to a tree--forest side solver.
We write $\beta_j$ for the exponential base of the solver
obtained after $j$ applications of the balanced-center
framework, with $\beta_0=2.3391$ for the auxiliary solver.
In the structural arguments, $E$ denotes a critical ee-set
for the fixed target $G$ supplied by
Lemma~\ref{lem:target-witness}, with $|E|=r$.
We use the ordering of its edges given by that lemma.
Finally, $d$ denotes the standard rSPR distance of a
$\rho$-augmented tree pair.

\begin{center}
\small
\begin{tabular}{p{0.16\linewidth}p{0.72\linewidth}}
\hline
Symbol & Meaning in this paper\\
\hline
$k$
  & supplied refined budget of the current instance
    above $\sigma_{\max}$\\
$\op_3$
  & optimum refined gap of the current instance,
    equal to $m(F_1,F_2)-\sigma_{\max}(F_1,F_2)$\\
$r$
  & refined gap of the fixed target $G$,
    equal to $\ord(G)-\sigma_{\max}(F_1,F_2)$\\
$b$
  & temporary budget used during the two
    binary-separation stages\\
$\kappa$
  & additional-cut budget passed to a tree--forest
    side solver\\
$\beta_j$
  & exponential base after $j$ applications of the
    balanced-center framework\\
$E$
  & critical ee-set of size $r$ for the fixed target,
    used with the ordering supplied by
    Lemma~\ref{lem:target-witness}\\
$d$
  & standard rSPR distance of a $\rho$-augmented
    tree pair\\
\hline
\end{tabular}
\end{center}

\subsection{Auxiliary Tree--Forest Solver}
\label{sec:auxiliary-ls-solver}

We use the split-or-decompose algorithm of Mestel et
al.~\cite{mestel2026split} as an auxiliary tree--forest solver.
Given an instance $(T,F)$ with nonnegative additional-cut budget
$\kappa$, it decides whether an \af of order at most
$\ord(F)+\kappa$ exists in $O^*(2.3391^\kappa)$ time.
We define the optimization solver \textsc{Alg-MAF$_M$} by running
this algorithm with successive budgets $j=0,1,\ldots,\kappa$
and returning the forest found at the first successful call.
The first successful budget is
$\op_2(T,F)=m(T,F)-\ord(F)$, so the returned forest is an \maf.
If all calls fail, the solver returns \FAIL.
The geometric sum over these calls gives the total running time
$\sum_{j=0}^{\min\{\op_2(T,F),\kappa\}}
    O^*(2.3391^j)
  =O^*(2.3391^{\min\{\op_2(T,F),\kappa\}})$.

If the label set is partitioned into two parts and no
component of either input contains labels from both parts,
the two restricted pairs can be solved independently.
The following lemma gives the corresponding additivity
of the optimum order and $\sigma_{\max}$.

\begin{lemma}
\label{lem:independent-decomposition}
Let $F_1$ and $F_2$ be two $X$-forests, and let
$X=X'\mathbin{\dot\cup}X''$, where $X'$ and $X''$ are nonempty.
Suppose that every component of $F_1$ and every component of $F_2$
has its label set contained entirely in either $X'$ or $X''$.
Let $F'$ be an \maf for $(F_1[X'],F_2[X'])$, and let $F''$ be an
\maf for $(F_1[X''],F_2[X''])$.  Then $F'\cup F''$ is an \maf for
$(F_1,F_2)$.  Conversely, if $F^*$ is an \maf for $(F_1,F_2)$, then
$F^*[X']$ and $F^*[X'']$ are \mafs for their respective restricted
pairs.
Consequently,
$m(F_1,F_2) = m(F_1[X'],F_2[X']) + m(F_1[X''],F_2[X''])$ and 
$\sigma_{\max}(F_1,F_2) =
\sigma_{\max}(F_1[X'],F_2[X'])
+ \sigma_{\max}(F_1[X''],F_2[X''])$.
\end{lemma}

\begin{proof}
By assumption, no component of either $F_1$ or $F_2$ contains labels
from both $X'$ and $X''$.  Hence every component of an \af for
$(F_1,F_2)$ is contained entirely in one of the two sets.  Therefore
any \af $F$ for $(F_1,F_2)$ decomposes as
$F=F[X']\cup F[X'']$,
where the two restrictions are \afs for
$(F_1[X'],F_2[X'])$ and $(F_1[X''],F_2[X''])$, respectively.
Conversely, the union of \afs for the two restricted
pairs is an \af for $(F_1,F_2)$.

It follows immediately that the union of two local \mafs is globally
optimal: otherwise, restricting a smaller global \af to $X'$ and
$X''$ would improve at least one of the two local solutions.
Conversely, if $F^*$ is a global \maf and one of its restrictions were
not locally optimal, replacing that restriction by a smaller \af would
produce a smaller global \af.  Hence
$m(F_1,F_2)
  =
  m(F_1[X'],F_2[X'])
  +
  m(F_1[X''],F_2[X''])$.

Every block of $\mathcal P(F_1,F_2)$ lies entirely in
either $X'$ or $X''$, since no component of either input
contains labels from both sides.
For each $Y\in\{X',X''\}$, restricting the inputs to $Y$
keeps exactly the components whose label sets are contained
in $Y$. Thus the blocks of $\mathcal P(F_1,F_2)$ contained
in $Y$ are exactly the blocks of
$\mathcal P(F_1[Y],F_2[Y])$.
Counting the blocks on the two sides and applying
Lemma~\ref{maxss} gives
$\sigma_{\max}(F_1,F_2) = \sigma_{\max}(F_1[X'],F_2[X'])
  + \sigma_{\max}(F_1[X''],F_2[X''])$.
\end{proof}

\section{Reduction Structure and Critical Edge-Deletion Sets}
\label{sec:Correctness}

This section develops the structural foundation used by the balanced
decomposition.  We first recall Reduction Rule~0 of Shi et
al.~\cite{shi2018jcss} and the associated greedy reduction process.
We then prove the main structural statement of the section: every
fixed target admits a critical ee-set whose edges can be ordered so
that each successive deletion increases $\sigma_{\max}$ by exactly
one. When the target is a \maf, the number of selected edges is
$\op_3$, and their locations will later be balanced inside the tree.

\subsection{Reduction Rule~0 and the greedy process}

The following rule is stated for an $X$-forest pair $(F_1,F_2)$ and
applies symmetrically after exchanging the two forests.

\medskip
\noindent {\it Reduction Rule~0.}
Let $e$ be an essential edge of $F_1$, and suppose that deleting $e$
splits a component $C_1$ into components $C_1'$ and $C_1''$.  If there
is a collection $\mathcal S_2$ of components of $F_2$ such that
$L(C_1')=L(C_1)\cap\bigl(\bigcup_{C\in\mathcal S_2}L(C)\bigr)$, then delete
$e$.

\medskip
An instance is \emph{$R_0$-reduced} if Reduction Rule~0 is
inapplicable to either forest with respect to the other.
Every edge deleted by Reduction Rule~0 is essential: its deletion
separates a current component into two nonempty labeled components.

\begin{lemma}[\cite{shi2018jcss}]
\label{jianer-1}
An edge deleted by Reduction Rule~0 is contained in no \af of the current pair.  Consequently, Reduction Rule~0 preserves
the set of \afs.
\end{lemma}

\begin{lemma}
\label{jianer-4}
Reduction Rule~0 preserves the component-pair partition
$\mathcal P(F_1,F_2)$ and hence $\sigma_{\max}(F_1,F_2)$.
\end{lemma}

\begin{proof}
Recall from Lemma~\ref{maxss} that $\sigma_{\max}(F_1,F_2)$ is the
number of blocks in the partition that places two labels in the same
block exactly when they lie in the same component of both forests.
Suppose that Reduction Rule~0 deletes $e$ from a component $C_1$ of
$F_1$ and produces sides $C_1'$ and $C_1''$.  By the condition of the
rule, the two sides are separated by a union of entire components of
$F_2$.  Hence two labels that lie in the same component of $F_2$
cannot be separated into $C_1'$ and $C_1''$.  Deleting $e$ therefore
does not split or merge any block of $\mathcal P(F_1,F_2)$.
Forced contraction changes no component label set. The partition
and its cardinality are therefore unchanged.
\end{proof}

\begin{lemma}
\label{jianer-2}
Let $F_1,F_2$ be $X$-forests, and let $F_i'$ be an
$X$-subforest of $F_i$ for $i\in\{1,2\}$.
Suppose Reduction Rule~0 applies to an edge $e$ in a component
$C$ of $F_1$ with respect to $F_2$, and let $U,V$ be the label
sets of the two components of $C\setminus\{e\}$.
Then exactly one of the following holds.
\begin{enumerate}[(i)]
  \item No component of $F_1'$ has labels in both $U$ and $V$,
        and $F_1'$ is an $X$-subforest of $F_1\setminus\{e\}$.
  \item A unique component $D$ of $F_1'$ has labels in both
        $U$ and $V$. It contains an edge $\bar e$ whose
        corresponding path in $F_1$ contains $e$.
        Reduction Rule~0 applies to $\bar e$ with respect to
        $F_2'$, and deleting $\bar e$ splits $D$ into components
        with label sets $L(D)\cap U$ and $L(D)\cap V$.
\end{enumerate}
The symmetric statement holds with the two input forests
interchanged.
\end{lemma}

\begin{proof}
Consider the minimal representations of the components of
$F_1'$ in $F_1$.
If no component has labels in both $U$ and $V$, all these
representations avoid $e$.
They therefore also realize $F_1'$ as an $X$-subforest of
$F_1\setminus\{e\}$, proving~(i).

Otherwise, let $D$ be a component with labels in both sets.
Its representation contains $e$.
Since distinct component representations are vertex-disjoint,
$D$ is the unique component with this property.
Reducing this representation by forced contraction produces $D$.
Let $\bar e$ be the edge of $D$ whose corresponding path
contains $e$.
Deleting $\bar e$ separates the nonempty label sets
$L(D)\cap U$ and $L(D)\cap V$.

It remains to verify Reduction Rule~0 for $\bar e$.
Since the rule applies to $e$, one side, say $U$, satisfies
$U=L(C)\cap\bigcup_{B\in\mathcal B}L(B)$ for a collection
$\mathcal B$ of components of $F_2$.
Let $\mathcal B'$ consist of the components of $F_2'$ whose
label sets are contained in those of components in $\mathcal B$.
The component partition of $F_2'$ refines that of $F_2$, and
$L(D)\subseteq L(C)$. Hence
\[
  L(D)\cap U
  =L(D)\cap\bigcup_{B\in\mathcal B}L(B)
  =L(D)\cap\bigcup_{B'\in\mathcal B'}L(B').
\]
Thus Reduction Rule~0 applies to $\bar e$ with respect to
$F_2'$, proving~(ii).
The two cases are mutually exclusive and exhaustive.
Exchanging the two inputs gives the symmetric statement.
\end{proof}

\begin{lemma}
\label{lem:one-step-target-separation}
Let $(F_1,F_2)$ be an $R_0$-reduced forest pair, and let
$G$ be an \af for this pair. Suppose that a component $A$ of $F_1$
contains the representations of at least two components of $G$.
Then there is an essential edge $e$ of $A$ such that
\begin{enumerate}[(i)]
  \item $e$ lies outside the representation of every component of
        $G$ in $F_1$, so $G$ remains an \af for
        $(F_1\setminus\{e\},F_2)$.
  \item $\sigma_{\max}(F_1\setminus\{e\},F_2)
        =\sigma_{\max}(F_1,F_2)+1$.
\end{enumerate}
\end{lemma}

\begin{proof}
Contract the pairwise vertex-disjoint representations of the
components of $G$ contained in $A$ into blocks, and let $K$ be
the minimal subtree connecting these blocks.
Since there are at least two blocks, $K$ has at least two leaves.
By minimality, every leaf of $K$ is a block.
Choose a leaf block corresponding to a component $C$ of $G$,
and let $e$ be the original edge corresponding to the unique
edge of $K$ incident to this block.

Every label of $A$ belongs to one of the contracted blocks.
Deleting $e$ therefore splits $A$ into two components
$A_C$ and $A_{\mathrm{rest}}$ with
$L(A_C)=L(C)$ and $L(A_{\mathrm{rest}})=L(A)\setminus L(C)$.
Both label sets are nonempty, so $e$ is essential.
Moreover, $e$ lies outside every target representation in $F_1$,
so $G$ remains an \af for $(F_1\setminus\{e\},F_2)$.
This proves~(i).

Let $D$ be the component of $F_2$ containing the representation
of $C$.
Since $L(A_C)=L(C)\subseteq L(D)$, only $D$ can have labels on
both sides of the cut.
If $D$ had no labels in $A_{\mathrm{rest}}$, then
$L(A)\cap L(D)=L(C)=L(A_C)$.
The single component $D$ would therefore satisfy the condition
of Reduction Rule~0 for $e$, contradicting the assumption that
$(F_1,F_2)$ is $R_0$-reduced.
Thus $D$ is the unique component of $F_2$ with labels on both
sides of the cut in $A$.
Deleting $e$ splits the block $L(A)\cap L(D)$ of the component-pair
partition into two nonempty blocks and leaves every other block
unchanged.
By Lemma~\ref{maxss}, $\sigma_{\max}$ equals the number of blocks,
so $\sigma_{\max}(F_1\setminus\{e\},F_2)
  =\sigma_{\max}(F_1,F_2)+1$.
This proves~(ii).
\end{proof}

\begin{corollary}
\label{cor:reduced-target-equality}
Let $(F_1,F_2)$ be $R_0$-reduced, and let $G$ be an \af for
this pair. If $\sigma_{\max}(F_1,F_2)=\ord(G)$, then
$(F_1,F_2)=(G,G)$ up to forced contraction.
\end{corollary}

\begin{proof}
If a component of either input contained two components of $G$,
Lemma~\ref{lem:one-step-target-separation} would give a cut that
preserves $G$ and increases $\sigma_{\max}$ to $\ord(G)+1$.
This contradicts Lemma~\ref{lem:j-new1}.
Since $G$ covers the full label set, every input component contains
at least one component of $G$. Hence it contains exactly one and
has the same label set as that component. Since $G$ is an \af,
the corresponding input and target components are isomorphic
after forced contraction. Thus $(F_1,F_2)=(G,G)$ up to forced
contraction.
\end{proof}

A \emph{run of \textsc{Greedy}} on an $X$-forest pair $(F_1,F_2)$
is a sequence of applications of Reduction Rule~0 in which, at each
step, an arbitrary applicable edge in either forest is chosen and the
rule is applied to that edge.  The run terminates when the current pair
is $R_0$-reduced, and the resulting pair is called its {\it terminal pair}.
Every run has polynomial length, since each application deletes an
edge, and each application of Reduction Rule~0 can be identified and
performed in polynomial time.

\begin{lemma}
\label{jianer-3}
If some sequence of Reduction Rule~0 applications transforms
$(F_1,F_2)$ into $(F^*,F^*)$, then $F^*$ is the unique \maf for
$(F_1,F_2)$, and every run of \textsc{Greedy} on $(F_1,F_2)$ has
terminal pair $(F^*,F^*)$.
\end{lemma}

\begin{proof}
Every application of Reduction Rule~0 preserves the set of
\afs. The assumed reduction to $(F^*,F^*)$
therefore shows that $F^*$ is the unique \maf, up to forced
contraction: every \af of that pair refines
$F^*$, and equality of orders forces equality of the forests.

Let $(A,B)$ be the terminal pair of any run of \textsc{Greedy}.
The forest $F^*$ remains an \af for this pair. Since the rule
also preserves $\sigma_{\max}$, comparison with the assumed
reduction sequence gives
$\sigma_{\max}(A,B) =\sigma_{\max}(F_1,F_2) =\sigma_{\max}(F^*,F^*)
  =\ord(F^*)$.
Corollary~\ref{cor:reduced-target-equality} now gives
$(A,B)=(F^*,F^*)$.
\end{proof}

\subsection{Critical ee-sets}

The balanced decomposition requires us to represent the
refined gap by edge locations in one input.
An edge set that cuts this input directly into the target
may be too large, because some separations are already
forced by the other input.
For example, suppose that $F$ is an $X$-subforest of $T$
and the target is $G=F$.
The refined gap for the pair $(T,F)$ is zero, whereas cutting $T$ directly
into $F$ requires $\ord(F)-1$ edges.
These cuts can instead be made by Reduction Rule~0.
We therefore consider edge sets whose deletion allows
the greedy reductions to recover the target.
We call such sets critical.
The definition below specifies this recovery condition;
Lemma~\ref{lem:target-witness} then shows that every target
admits a critical ee-set whose size equals its refined gap.

\begin{definition}
\normalfont
An ee-set $E_1$ of $F_1$ is \emph{critical} for $(F_1,F_2)$ with
respect to an \af $G$ if
\textsc{Greedy}$(F_1\setminus E_1,F_2)$ returns $(G,G)$.
Critical ee-sets of $F_2$ are defined symmetrically.
\end{definition}

The following lemma gives a critical ee-set whose size equals
the target's refined gap, with an ordering in which each
successive edge deletion increases $\sigma_{\max}$ by one.
The target can be any \af.

\begin{lemma}
\label{lem:target-witness}
Let $G$ be any \af for $(F_1,F_2)$, and let
$r=\ord(G)-\sigma_{\max}(F_1,F_2)$.  There is a critical ee-set
$E=\{e_1,\ldots,e_r\}$ of $F_1$ with respect to $G$ that admits an
ordering $e_1,\ldots,e_r$ such that, for every $1\leq i\leq r$,
$\sigma_{\max}(F_1\setminus\{e_1,\ldots,e_i\},F_2) =\sigma_{\max}(F_1,F_2)+i$.
Given $F_1,F_2$, and $G$, such a critical ee-set and ordering can be
constructed in polynomial time.  The symmetric statement holds for
$F_2$.
\end{lemma}

\begin{proof}
Let $\sigma_0=\sigma_{\max}(F_1,F_2)$.
Since $G$ is an \af, we have $\sigma_0\leq\ord(G)$, so $r\geq0$.
Throughout the construction, we keep track of the original edges
represented by each current edge.
When forced contraction suppresses a path into a single edge,
deleting the resulting edge can be implemented by deleting a
surviving original edge on that path.
We use such an original edge as the representative of each
selected cut. The same convention records every Reduction
Rule~0 cut in either input.

First apply Reduction Rule~0 exhaustively to $(F_1,F_2)$,
obtaining an $R_0$-reduced pair $(A_0,B_0)$.
By Lemmas~\ref{jianer-1} and~\ref{jianer-4}, $G$ remains an
\af and
$\sigma_{\max}(A_0,B_0)=\sigma_0$.
We construct $R_0$-reduced pairs $(A_j,B_j)$ and distinct original
edges $e_1,\ldots,e_j$ such that
$G$ is an \af for $(A_j,B_j)$, and 
$\sigma_{\max}(A_j,B_j)=\sigma_0+j$.

Suppose that $j<r$.
Some component of $A_j$ must contain the representations of at
least two components of $G$.
Indeed, otherwise every component of $A_j$ would contain exactly
one target component, giving $\ord(A_j)=\ord(G)$.
This would contradict
$\ord(A_j)
\leq \sigma_{\max}(A_j,B_j)
=\sigma_0+j
<\ord(G)$.
Lemma~\ref{lem:one-step-target-separation} therefore supplies an
essential edge $f_{j+1}$ of $A_j$ whose deletion preserves $G$
and increases $\sigma_{\max}$ by exactly one.
Record an original-edge representative $e_{j+1}$ of this cut,
delete $f_{j+1}$, and apply Reduction Rule~0 exhaustively.
The resulting pair $(A_{j+1},B_{j+1})$ again satisfies the
invariant, because these reductions preserve both $G$ and
$\sigma_{\max}$. After $r$ steps, we obtain the recorded set
$E=\{e_1,\ldots,e_r\}$.

We next verify the order increase for each prefix in the
original forest. Let $E_j=\{e_1,\ldots,e_j\}$, with
$E_0=\varnothing$. Before the $(j+1)$-st selected cut, $A_j$
is obtained from $F_1$ by the selected cuts in $E_j$, further
Reduction Rule~0 cuts, and forced contraction. In the original
graph with only $E_j$ deleted, before contraction, the path
representing $f_{j+1}$ is therefore still present. Each side of
$e_{j+1}$ contains the labels from the corresponding nonempty
side of $f_{j+1}$. Deleting $e_{j+1}$ separates these labels.
Consequently,
$\ord(F_1\setminus E_{j+1})
  =\ord(F_1\setminus E_j)+1$ for $0\leq j<r$.
This statement concerns deletion at the recorded original
location; it does not require $e_{j+1}$ itself to survive as an
edge after contraction. Summing the order increases proves that
$E$ is an ee-set.

Fix $0\leq i\leq r$. We claim that the recorded Reduction Rule~0 steps transform
$(F_1\setminus E_i,F_2)$ into $(A_i,B_i)$.
To see this, replay the initial reductions and the reductions
following the first $i$ selected cuts, in their original order.
The selected cuts themselves have already been made.
At each replayed step, the current pair is a subforest of the
pair on which that reduction was originally applied: the extra
selected cuts have only refined its components.
Lemma~\ref{jianer-2} supplies an applicable current representative
whenever a component still contains labels from both sides of
the recorded cut. We apply the rule to that representative.
We skip the cut only if no current component meets both sides;
in that case the lemma shows that it is redundant.

This replay makes the same selected and reduction cuts as the
construction through stage $i$, with only their order changed.
Since edge deletions commute before forced contraction, and
skipped cuts do not change the resulting forest, the replay
ends at $(A_i,B_i)$ up to forced contraction.
Reduction Rule~0 preserves $\sigma_{\max}$, so
$\sigma_{\max}(F_1\setminus E_i,F_2) =
\sigma_{\max}(A_i,B_i)
=
\sigma_0+i$.
This proves all the required prefix equalities.

At the last stage,
$\sigma_{\max}(A_r,B_r)=\sigma_0+r=\ord(G)$.
Corollary~\ref{cor:reduced-target-equality} therefore identifies
$(A_r,B_r)$ with $(G,G)$.
Taking $i=r$ in the replay argument gives a sequence of
Reduction Rule~0 applications transforming
$(F_1\setminus E,F_2)$ into $(G,G)$.
By Lemma~\ref{jianer-3}, every run of
\textsc{Greedy}$(F_1\setminus E,F_2)$ returns $(G,G)$.
Hence $E$ is critical with respect to $G$.

The target representations, the cuts supplied by
Lemma~\ref{lem:one-step-target-separation}, and the applicable
Reduction Rule~0 steps can all be found in polynomial time.
There are $r$ selected cuts, every reduction sequence has
polynomial length, and original-edge representatives can be
maintained through forced contractions in polynomial time.
The entire construction therefore takes polynomial time.
The construction for $F_2$ is symmetric.
\end{proof}

In particular, choosing $G$ to be an \maf gives a critical
ee-set of size $\op_3(F_1,F_2)$ with the same prefix property.
Thus the optimum refined gap is represented by edge locations
in either input forest. The next section uses these locations
to obtain a balanced decomposition.

\section{Balanced Centers and Side Parameters}
\label{sec:balanced-centers}

We now specialize to tree--forest pairs $(T,F)$, where $T$
is an $X$-tree and $F$ is an $X$-forest. Fix a target \af $G$
for $(T,F)$, and let $E$ be the critical ee-set for $G$ with
the ordering supplied by Lemma~\ref{lem:target-witness}.
When $|E|\geq 3$, we first show that there is a nonroot internal
vertex of $T$ such that each of its three sides contains at
most half of the edges in $E$.
For a side with label set $Y$, we then show that the refined
gap of $G[Y]$ for $(T[Y],F[Y])$ is at most the number of edges
of $E$ lying in that side. The proof follows the changes in the
component-pair partition restricted to $Y$ as the edges of $E$
are deleted in order. Only a critical edge in the side can
increase the number of blocks, and each such edge increases
it by at most one. Thus, at the chosen vertex, the refined gap
of each restricted target is at most $\lfloor |E|/2\rfloor$,
which also bounds the refined parameter of each side instance.

\subsection{Balanced centers}

For a vertex $v$ of $T$, let $T-v$ denote the graph obtained
by deleting $v$ and its incident edges, without applying
forced contraction.
If $c$ is a nonroot internal vertex, then $T-c$ has three
connected components.
Let $Q_0(c)$ be the component containing the root of $T$, and let
$Q_1(c),Q_2(c)$ be the components containing the two children
of $c$, in an arbitrary but fixed order.
We call these components the three \emph{sides} at $c$.

For $i\in\{0,1,2\}$, let $M_i(c,E)=|E\cap E(Q_i(c))|$ for an arbitrary edge set $E\subseteq E(T)$.
Thus $M_i(c,E)$ counts the edges of $E$ lying entirely in the
$i$-th side.
Every edge of $E$ not incident with $c$ is counted on exactly
one side, while edges incident with $c$ are not counted.
These counts refer to original edges, before forced contraction is performed.
A nonroot internal vertex $c$ is a \emph{balanced center}
for $E$ if $M_i(c,E)\leq |E|/2$ for every $i\in\{0,1,2\}$.
The following lemma establishes the existence of such a vertex
for any edge set of size at least three.

\begin{lemma}
\label{lem:balanced-center-generic}
Let $T$ be an $X$-tree with at least three leaves,
and let $E\subseteq E(T)$ with $|E|=m\geq3$.
There is a nonroot internal vertex $c$ such that
$M_i(c,E)\leq m/2$ for every $i\in\{0,1,2\}$.
\end{lemma}

\begin{proof}
Assign weight zero to the root of $T$.
For each nonroot vertex $v$, let $p(v)$ be its parent and set $w(v)= 1$ if the edge $(p(v),v) \in E$ and $w(v)= 0$ otherwise.
Each edge of $E$ therefore contributes one unit of weight
at its child endpoint, and the total vertex weight is $m$.
For a subgraph $Q$ of $T$, write $w(Q)$ for its total
vertex weight.

If an edge is counted by $M_i(c,E)$, then both endpoints lie in
$Q_i(c)$, and its child endpoint contributes one unit to
$w(Q_i(c))$.
Different edges have different child endpoints, so
$M_i(c,E) \leq w(Q_i(c))$ for each $i \in \{0,1,2\}$.
It suffices, therefore, to find a nonroot internal vertex
$c$ for which all three side weights are at most $m/2$.

We first find a weighted centroid of $T$.
Choose a vertex $z$ minimizing
$\Phi(z)=\sum_{v\in V(T)} w(v) \cdot \operatorname{dist}_T(z,v)$,
where $\operatorname{dist}_T(z,v)$ is the distance between $z$ and $v$ (i.e., the number of edges on the shortest path connecting $z$ and $v$) in the underlying undirected tree.
For any neighbor $u$ of $z$, let $C$ be the component of
$T-z$ containing $u$.
Since $z$ minimizes $\Phi$, we have $\Phi(u)\geq\Phi(z)$.
Moving from $z$ to $u$ decreases the total weighted distance
by $w(C)$ for vertices in $C$ and increases it by $m-w(C)$
for vertices outside $C$.
Therefore,
\[
0\leq\Phi(u)-\Phi(z)
=-w(C)+(m-w(C))
=m-2w(C),
\]
and hence $w(C)\leq m/2$.
Therefore every component of $T-z$ has weight at most $m/2$.

The vertex $z$ cannot be a leaf.
Indeed, a leaf has weight at most one, so its removal leaves
a connected component of weight at least
$m-1>m/2$, since $m\geq3$.
If $z$ is a nonroot internal vertex, take $c=z$.
The centroid property immediately gives
$M_i(c,E)\leq w(Q_i(c))\leq m/2$
for each $i \in \{0,1,2\}$.

It remains to consider the case where $z$ is the root.
Since $T$ has at least three leaves, the root has an internal
child $c$.
Let $u$ be its other child, and let $T_c,T_u$ be the
subtrees rooted at $c,u$, respectively.
These are the two components of $T-z$, so
$w(T_c)\leq m/2$ and $w(T_u)\leq m/2$.
The root side $Q_0(c)$ consists of $z$ and $T_u$.
Since $w(z)=0$, $M_0(c,E)\leq w(Q_0(c))=w(T_u)\leq m/2$.
The child sides $Q_1(c)$ and $Q_2(c)$ are contained in $T_c$.
By nonnegativity of the weights,
$M_i(c,E)\leq w(Q_i(c))\leq w(T_c)\leq m/2$ for each $i \in \{1,2\}$.
Thus $c$ is a balanced center in both cases.
\end{proof}

For a fixed target $G$, Lemma~\ref{lem:target-witness}
supplies a critical ee-set $E$ of size
$r=\ord(G)-\sigma_{\max}(T,F)$.
When $r\geq3$, Lemma~\ref{lem:balanced-center-generic} guarantees
a balanced center for this set.
The algorithm does not know $G$ or $E$.
Instead, it enumerates all nonroot internal vertices of $T$,
one of which has the required balance for the fixed target.
The cases $r\leq2$ are handled by the auxiliary solver.

Additionally, the three sides can be separated by deleting only
the two child edges of $c$.
Let $X_i=L(Q_i(c))$ for $i\in\{0,1,2\}$, and let $f_1,f_2$
be the edges joining $c$ to its two children.
Deleting $f_1,f_2$ leaves the two child sides and a root-side
component consisting of $Q_0(c)$ together with $c$ and its
parent edge.
The vertex $c$ is then unlabeled and has undirected degree
one, so forced contraction removes it.
After forced contraction, the three components
are $T[X_0],T[X_1],T[X_2]$.
Their label sets are precisely $X_0,X_1,X_2$, respectively.

\subsection{Side-parameter bounds}

The count $M_i(c,E)$ records the original edges of $E$
lying in $Q_i(c)$.
We now relate this count to the refined parameter of
$(T[X_i],F[X_i])$.
To account for restriction and forced contraction, we use the
component-pair partition defined in Section~\ref{sec:problem-formulations}.
For any two $X$-forests $H_1,H_2$ and any nonempty $Y\subseteq X$,
restriction acts on their component-pair partition as follows:
$\mathcal P(H_1[Y],H_2[Y])
= \{P\cap Y: P \in \mathcal P(H_1,H_2),\;
P \cap Y \neq \varnothing \}$.

\begin{lemma}
\label{lem:restricted-critical-replay}
Let $G$ be an \af for a tree--forest pair $(T,F)$, and let
$E\subseteq E(T)$ be an ee-set critical for $(T,F)$ with
respect to $G$.
Then, for every nonempty $Y\subseteq X$,
$\sigma_{\max}((T\setminus E)[Y],F[Y]) = \ord(G[Y])$.
\end{lemma}

\begin{proof}
Since $E$ is critical with respect to $G$, a sequence of
Reduction Rule~0 applications transforms
$(T\setminus E,F)$ into $(G,G)$.
By Lemma~\ref{jianer-4},
$\mathcal P(T\setminus E,F)
=
\bigl\{
L(C): C\text{ is a component of }G
\bigr\}$.
Restricting both sides of this identity to $Y$ gives
$\mathcal P((T\setminus E)[Y],F[Y])
=\{L(C)\cap Y: C\text{ is a component of }G,\;
L(C)\cap Y\neq\varnothing\}$.
The sets on the right are precisely the component label
sets of $G[Y]$.
Their number is $\ord(G[Y])$, proving the claim.
\end{proof}

The next lemma uses the prefix property of the critical
ee-set to bound how much $\sigma_{\max}$ can increase
after restriction to a side.

\begin{lemma}
\label{lem:side-parameter}
Let $G$ be an \af for a tree--forest pair $(T,F)$, and let
$E=\{e_1,\ldots,e_r\}\subseteq E(T)$ be an ee-set critical
for $(T,F)$ with respect to $G$.
Suppose that the ordering satisfies
$\sigma_{\max}(T\setminus\{e_1,\ldots,e_j\},F)
= \sigma_{\max}(T,F)+j$ for each $1 \leq j\leq r$.
Let $Q$ be a component of $T-e$ for some edge $e$ of $T$,
and let $Y=L(Q)$ and $E_Y=E\cap E(Q)$.
Then $\ord(G[Y])-\sigma_{\max}(T[Y],F[Y]) \leq |E_Y|$,
and consequently $\op_3(T[Y],F[Y])\leq |E_Y|$.
\end{lemma}

\begin{proof}
For $0\leq j\leq r$, let $E_j=\{e_1,\ldots,e_j\}$ and $H_j=T\setminus E_j$,
where $E_0=\varnothing$.
Let $\mathcal P_j=\mathcal P(H_j,F)$, and 
$s_j=\sigma_{\max}(H_j[Y],F[Y])$.
Thus $s_j$ is the number of nonempty intersections of
blocks of $\mathcal P_j$ with $Y$.

Consider the deletion of the original edge $e_j$ after
the edges in $E_{j-1}$ have been deleted.
Before forced contraction, this deletion can split only
one component of the first forest.
Each block of $\mathcal P_{j-1}$ therefore either remains
unchanged or splits into two nonempty blocks.
By the prefix property, $|\mathcal P_j|-|\mathcal P_{j-1}|=1$.
Hence exactly one block $P$ splits, say into $P_1,P_2$;
all other blocks are unchanged.

After intersection with $Y$, this split increases the
number of nonempty blocks by one if both $P_1\cap Y$
and $P_2\cap Y$ are nonempty, and by zero otherwise.
It follows that $0\leq s_j-s_{j-1} \leq 1$.

Moreover, if $e_j\notin E_Y$, the increase is zero.
Indeed, $e_j$ does not belong to the connected subgraph
$Q$, so all labels in $Y$ lie in the same component of
$T-e_j$.
Deleting $e_j$, even after the preceding deletions, cannot
separate two labels of $Y$ that are still connected.
Thus the split of $P$ cannot have labels of $Y$ on both
sides.
Consequently, $s_j-s_{j-1}=0$ whenever $e_j\notin E_Y$.
Summing over the ordering gives
$s_r-s_0 = \sum_{j=1}^{r}(s_j-s_{j-1}) \leq |E_Y|$.

Initially,
$s_0=\sigma_{\max}(T[Y],F[Y])$.
By Lemma~\ref{lem:restricted-critical-replay},
$s_r=\ord(G[Y])$.
Thus $\ord(G[Y])-\sigma_{\max}(T[Y],F[Y]) \leq |E_Y|$.
Finally, $G[Y]$ is an \af for $(T[Y],F[Y])$, so
$\op_3(T[Y],F[Y]) \leq \ord(G[Y])-\sigma_{\max}(T[Y],F[Y]) \leq |E_Y|$.
\end{proof}

Combining this bound with the balanced-center property
gives the required bounds for the three side instances.

\begin{corollary}
\label{cor:balanced-side-parameter}
Let $G$ be an \af for $(T,F)$, and let $E$ be the critical
ee-set supplied by Lemma~\ref{lem:target-witness}, with its
guaranteed ordering.
Let $r=|E|\geq3$, let $c$ be a balanced center for $E$,
and let $X_i=L(Q_i(c))$ for $i\in\{0,1,2\}$.
Then, for every $i\in\{0,1,2\}$,
$\ord(G[X_i])-\sigma_{\max}(T[X_i],F[X_i])
\leq M_i(c,E)\leq\lfloor r/2\rfloor$,
and $\op_3(T[X_i],F[X_i])
\leq M_i(c,E)\leq\lfloor r/2\rfloor$.
In particular, if $r\leq k$, every side instance
has refined parameter at most $\lfloor k/2\rfloor$.
\end{corollary}

The preceding results guarantee that, for at least one
enumerated center, each side instance has refined parameter
at most $\lfloor r/2\rfloor$.
To use these bounds algorithmically, we still need to
separate $F$ according to the same label partition and
recover a global optimum from the side solutions.
This requires care because a component of the target
\af may contain labels from different sides.
The next two sections develop the separation and
reconstruction procedures.
Section~\ref{sec:balanced-center-algorithm} then combines
these procedures into the full algorithm and analyzes
its correctness and running time.

\section{Easy Binary Separation and Optional Reconstruction}
\label{sec:easy}

Let $(T,F)$ be a tree--forest pair on $X$, and let
$e=(w,v)$ be an edge of $T$, where $w$ is the parent of $v$.
Let $T_v$ and $T_w$ be the two connected components of $T-e$
containing $v$ and $w$, respectively.
Let $X_v=L(T_v)$ and $X_w=L(T_w)$.
Thus $X_v=L(v)$ is the descendant label set of $v$,
whereas $X_w=X\setminus X_v$ is the label set on the root side. 
The edge $e$ induces the label bipartition $(X_v,X_w)$ of $X$.
An $X$-forest $H$ \emph{admits} $(X_v,X_w)$ if every component of
$H$ has all of its labels in one side of the bipartition.  Equivalently,
$X_v$ is the union of the label sets of some components of $H$.  We
call the bipartition \emph{easy for $H$} if $H$ admits it or if there
is an edge $e'$ of $H$ such that $H\setminus\{e'\}$ admits it.  These
conditions can be tested in linear time.

We show how to realize an easy bipartition while retaining the
information needed for reconstruction.  The separated sides are kept
for further separation or side solving in the balanced-center
algorithm.  Parameter balance and the overall running-time analysis
are deferred to Section~\ref{sec:balanced-center-algorithm}.

\subsection{Direct separation}

If $F$ already admits $(X_v,X_w)$, then $X_v$ is a union of
component label sets of $F$, so Reduction Rule~0 applies to $e$ in
$T$.  Lemmas~\ref{jianer-1} and~\ref{jianer-4} give the following
lemma. No marker is needed in this case.

\begin{lemma}
\label{lem:direct-easy-separation}
If $F$ admits $(X_v,X_w)$, then $(T,F)$ and
$(T\setminus\{e\},F)$ have the same \afs and the same
value of $\sigma_{\max}$. Both forests in the latter pair admit
$(X_v,X_w)$. The two independent side instances are
$(T[X_v],F[X_v])$ and $(T[X_w],F[X_w])$.
\end{lemma}

\subsection{Local marker operation}

\begin{figure}[tbp]
\centering
\begin{subfigure}[b]{0.46\linewidth}
\centering
\begin{tikzpicture}[
  scale=.72,
  vertex/.style={circle,fill=black,inner sep=1.2pt},
  marker/.style={circle,draw,inner sep=1pt,font=\scriptsize}]
  \node[font=\scriptsize] at (0,1.8) {$F_T$};
  \node[vertex,label=above:{\scriptsize $w$}] (w) at (-.7,.9) {};
  \node[marker] (g) at (-.2,.2) {$\gamma$};
  \draw (w)--(g);
  \node[vertex] (pv) at (.9,.9) {};
  \node[vertex,label=below:{\scriptsize $v$}] (v) at (.55,.1) {};
  \node[marker] (d) at (1.25,.1) {$\delta$};
  \draw (pv)--(v) (pv)--(d);

  \node[font=\scriptsize] at (3.1,1.8) {$F_F$};
  \node[vertex,label=above:{\scriptsize $w'$}] (wp) at (2.4,.9) {};
  \node[marker] (gp) at (2.9,.2) {$\gamma$};
  \draw (wp)--(gp);
  \node[vertex] (pvp) at (4,.9) {};
  \node[vertex,label=below:{\scriptsize $v'$}] (vp) at (3.65,.1) {};
  \node[marker] (dp) at (4.35,.1) {$\delta$};
  \draw (pvp)--(vp) (pvp)--(dp);
\end{tikzpicture}
\caption{Parallel orientation.}
\end{subfigure}%
\hspace{3mm}%
\begin{subfigure}[b]{0.46\linewidth}
\centering
\begin{tikzpicture}[
  scale=.72,
  vertex/.style={circle,fill=black,inner sep=1.2pt},
  marker/.style={circle,draw,inner sep=1pt,font=\scriptsize}]
  \node[font=\scriptsize] at (0,1.8) {$F_T$};
  \node[vertex,label=above:{\scriptsize $w$}] (w) at (-.7,.9) {};
  \node[marker] (g) at (-.2,.2) {$\gamma$};
  \draw (w)--(g);
  \node[vertex] (pv) at (.9,.9) {};
  \node[vertex,label=below:{\scriptsize $v$}] (v) at (.55,.1) {};
  \node[marker] (d) at (1.25,.1) {$\delta$};
  \draw (pv)--(v) (pv)--(d);

  \node[font=\scriptsize] at (3.1,1.8) {$F_F$};
  \node[vertex,label=above:{\scriptsize $w'$}] (wp) at (2.4,.9) {};
  \node[marker] (dp) at (2.9,.2) {$\delta$};
  \draw (wp)--(dp);
  \node[vertex] (pvp) at (4,.9) {};
  \node[vertex,label=below:{\scriptsize $v'$}] (vp) at (3.65,.1) {};
  \node[marker] (gp) at (4.35,.1) {$\gamma$};
  \draw (pvp)--(vp) (pvp)--(gp);
\end{tikzpicture}
\caption{Opposite orientation.}
\end{subfigure}
\caption{Local schematic views of the two marker modifications.
Unchanged parent edges, sibling branches, and descendants of $v$
and $v'$ are omitted; the drawings are not complete forests.
The deleted connectors are not shown.  In Modification~(I), equal marker labels occupy corresponding
local positions.  In Modification~(II), the marker names are exchanged
in $F_F$ to compensate for the opposite orientations of $e$ and $e'$.}
\label{fig:Case 2.2}
\end{figure}
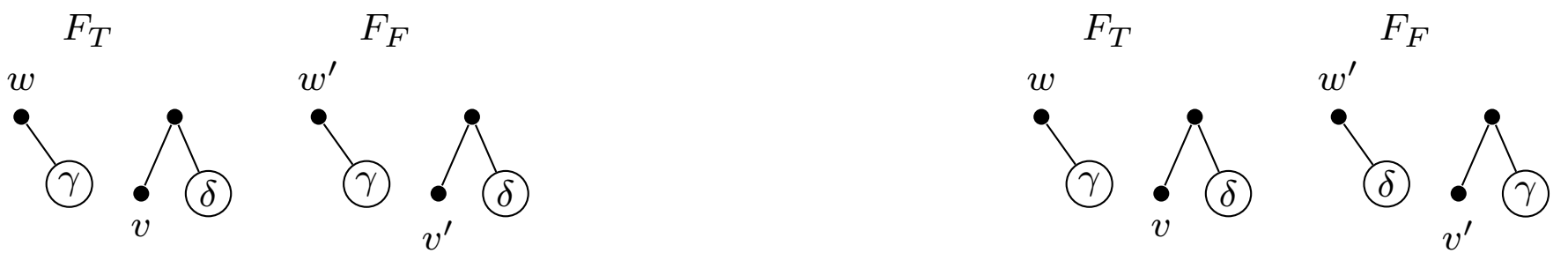

For separations that use markers, we define a local operation shared
by this section and Section~\ref{sec:difficult}.  Let $f=(w',v')$ be
an edge of $F$, where $w'$ is the parent of $v'$, and choose distinct
fresh labels $\delta,\gamma\notin X$.  Delete $e$ and $f$ without
applying forced contraction.  In $T-e$, introduce a new
parent of $v$ whose two children are $v$ and a marker leaf labeled
$\delta$, and attach a marker leaf labeled $\gamma$ as a child of
$w$.  Modify $F-f$ at $v'$ and $w'$ in the same way,
with marker labels determined by the orientation:
\begin{itemize}
  \item \emph{Parallel.} Make $v'$ and $\delta$ the two children
        of a new root, and attach $\gamma$ as a child of $w'$.
  \item \emph{Opposite.} Make $v'$ and $\gamma$ the two children
        of a new root, and attach $\delta$ as a child of $w'$.
\end{itemize}
In both orientations, the marker placement in the first input is
unchanged.  Apply forced contraction only after all marker leaves
have been inserted.  Store the ordered marker pair $(\delta,\gamma)$
and the orientation in a constant-size reconstruction record.
Figure~\ref{fig:Case 2.2} illustrates the two orientations.

The operation applies to any edge $f$ of $F$. In the one-edge
case below, deleting $f$ separates the labels in $X_v$ from
those in $X_w$ within the component containing $f$. The
orientation is determined by which of these two sets contains
the labels below $v'$.
In Section~\ref{sec:difficult}, the
algorithm tries each edge of $F$ in the parallel orientation.
For each choice, the two-color process searches for further
deletions to separate the two sides in the marked second forest.

\subsection{Separation exposed by one edge}

Suppose that $F$ does not admit $(X_v,X_w)$, but
$F\setminus\{e'\}$ does for an edge $e'=(w',v')$, where $w'$
is the parent of $v'$. Let $C_{e'}$ be the component of $F$
containing $e'$. Since deleting $e'$ affects only $C_{e'}$,
this is the only component with labels in both $X_v$ and $X_w$.
In $F-e'$, the component rooted at $v'$ has label set $L(v')$,
and the component containing $w'$ has the remaining labels of
$C_{e'}$. These two components lie on different sides of the
bipartition.

The marker operation on $T$ places $\delta$ with $X_v$ and
$\gamma$ with $X_w$. To match these assignments in $F$, the
component rooted at $v'$ must receive $\delta$ if its labels
lie in $X_v$ and $\gamma$ if they lie in $X_w$. We therefore
apply the local marker operation with $f=e'$ in the parallel
orientation (Modification~(I)) if $L(v')=L(C_{e'})\cap X_v$,
and in the opposite orientation (Modification~(II)) if
$L(v')=L(C_{e'})\cap X_w$. Write $F_T$ and $F_F$ for the
resulting forests, and let $X'=X\cup\{\delta,\gamma\}$,
$X'_v=X_v\cup\{\delta\}$, and $X'_w=X_w\cup\{\gamma\}$.
Both marked forests admit $(X'_v,X'_w)$.

We construct an \af for $(F_T,F_F)$ from a target \af $G$
for $(T,F)$. The construction depends on whether the
representation of $G$ in $T$ contains $e$.

\begin{lemma}
\label{lem:split-marked-target}
Let $G$ be an \af for $(T,F)$ with $e\in E_T(G)$, and let $C$
be the unique component whose representation in $T$ contains
$e$. Let $A=L(C)\cap X_v$ and $B=L(C)\cap X_w$. Both sets are
nonempty, and $C[A]$ is a pendant subtree of $C$. Define
$G_{X'}$ by replacing $C$ with the two trees below and retaining
all other components of $G$.
\begin{itemize}
  \item $C_\delta$ is obtained by adding a new root whose
        children are $C[A]$ and the marker leaf $\delta$;
  \item $C_\gamma$ is obtained from $C$ by replacing the
        pendant subtree $C[A]$ with the marker leaf $\gamma$.
\end{itemize}
Then $G_{X'}$ is an \af for $(F_T,F_F)$ of order $\ord(G)+1$,
and both markers belong to non-singleton components.
\end{lemma}

\begin{proof}
Deleting $e$ splits the minimal representation of $C$ in
$T$ into two parts with nonempty label sets $A$ and $B$.
The descendant part has labels $A$, so $e$ lies on the path
representing the edge entering $C[A]$. In $F$, deleting $e'$
separates the labels in $A$ from those in $B$, so the
representation of $C$ in $F$ must also contain $e'$.
In each input, cutting the representation at its connector
and inserting the markers gives two vertex-disjoint subgraphs,
with label sets $A\cup\{\delta\}$ and $B\cup\{\gamma\}$.
These subgraphs use only vertices of the original representation
of $C$ and fresh vertices. They are therefore disjoint from
the representations of all other components of $G$.

In $T$, the marker operation puts a new root above $v$ and
attaches $\delta$ to that root. It attaches $\gamma$ to $w$,
the parent endpoint of $e$ on the path entering $C[A]$.
Forced contraction reduces the two parts to $C_\delta$ and
$C_\gamma$, respectively. We now check the two orientations
in $F$.

\emph{Parallel orientation.}
The labels of $C$ below $e'$ are exactly $A$, so $e'$ lies
on the path representing the same edge of $C$ as $e$.
The marker operation acts on the same branch of $C$ and
places the markers on the same sides. Forced contraction
therefore gives $C_\delta$ and $C_\gamma$ in $F$ as well.

\emph{Opposite orientation.}
The labels of $C$ below $e'$ are exactly $B$, so $B$ is
also a descendant label set in $C$. The nonempty sets $A$ and
$B$ partition $L(C)$, and each lies below one of the root's
two children. To cover all of $L(C)$, they must lie below
different children and each must contain all the labels below
its child. Thus $C[A]$ and $C[B]$ are the two child subtrees
of the root, and $C_\gamma$ has root children $C[B]$ and
$\gamma$. The operation in $F$ puts $C[B]$ and $\gamma$
under a new root after contraction, giving $C_\gamma$.
In the other part, replacing the branch for $C[B]$ with
$\delta$ gives a root with children $C[A]$ and $\delta$.
This is $C_\delta$.

The disjoint representations constructed above therefore give
an \af $G_{X'}$ for $(F_T,F_F)$. Replacing $C$ by two trees
increases the order by one. Each new tree contains a nonempty
set of labels from $X$ as well as its marker.
\end{proof}

\begin{lemma}
\label{lem:singleton-marker-target}
Let $G$ be an \af for $(T,F)$ with $e\notin E_T(G)$, and let $G^\circ=G\cup\{\{\delta\},\{\gamma\}\}$. Then $G^\circ$
is an \af for $(F_T,F_F)$.
Moreover, $\ord(G^\circ)=\ord(G)+2$, and both markers
are singleton components.
\end{lemma}

\begin{proof}
Since the representation of $G$ in $T$ avoids $e$,
every component of $G$ has all its labels in $X_v$
or all in $X_w$.
Deleting $e'$ separates these two label sets within
$C_{e'}$, and all other components of $F$ already lie
on one side. Therefore the representation in $F$ of
every component of $G$ also avoids $e'$.
It follows that $G$ remains an \af after
the two connectors are deleted.

Insert the markers as specified by either modification,
but retain each marker as a singleton component in the
\af. The original component representations
remain unchanged, and each singleton marker is represented
only by its new labeled leaf. These representations are
pairwise disjoint in both modified inputs.
Hence $G^\circ$ is an \af for $(F_T,F_F)$.
Adding the two singleton components increases the order
by two.
\end{proof}

The direct and marked constructions are used by the binary
separation procedure in Section~\ref{subsec:deferred-binary}.
Before assembling that procedure, we describe how to reconstruct
an \af from its side solutions.

\subsection{Optional reconstruction}
\label{subsec:optional-reconstruction}

After separation, the two side instances can be solved
independently. A component of an \af for the original pair
may contain labels from both sides, so we use the marker
leaves to recover a connection when the returned forests
support it. Write $H'$ for the marked forest to which we
apply the rule and $H$ for the reconstructed forest. The
record identifies the two markers and the orientation.
Reconstruction uses the marker positions in $H'$, where
forced contraction may have removed the original attachment
vertices.

\begin{lemma}[Local reconstruction]
\label{lem:maf combination for case 2.2}
Let $(T,F)$ be a tree--forest pair on $X$, and let $(F_T,F_F)$
be obtained by applying the local marker operation to edges
$e$ of $T$ and $f$ of $F$, with fresh markers $\delta,\gamma$.
Let $H'$ be any \af for $(F_T,F_F)$, and let $H_\delta$ and
$H_\gamma$ be the components of $H'$ containing $\delta$ and
$\gamma$, respectively. The following rule gives an \af $H$
for $(T,F)$.
\begin{enumerate}[(i)]
  \item If both marker components are non-singletons, write
        $H_\delta^-$ and $H_\gamma^-$ for their restrictions to
        $L(H_\delta)\setminus\{\delta\}$ and
        $L(H_\gamma)\setminus\{\gamma\}$, respectively.
        Replace $H_\delta$ and $H_\gamma$ by one tree as follows.
        \begin{itemize}
          \item \emph{Parallel.} In $H_\gamma$, replace the
                leaf $\gamma$ with $H_\delta^-$ by making
                its root a child of the former parent of $\gamma$.
          \item \emph{Opposite.} Make the roots of $H_\delta^-$
                and $H_\gamma^-$ the two children of a new root.
        \end{itemize}
        In either orientation, $\ord(H)=\ord(H')-1$.
  \item If at least one marker component is a singleton, set
        $H=H'[X]$. Its order is $\ord(H')-1$ if exactly one
        marker is a singleton and $\ord(H')-2$ if both are.
\end{enumerate}
Every component containing neither marker is kept. The label
set $X$ may include markers from earlier operations.
\end{lemma}

\begin{proof}
The markers lie in separate components of $F_T$, so
$H_\delta$ and $H_\gamma$ are distinct.

For~(i), write $e=(w,v)$ and $f=(w',v')$, with $v$ and $v'$
the child endpoints. Consider the representations of
$H_\delta$ and $H_\gamma$ before the forced contractions
following marker insertion. Each representation contains a
path from its marker to a label in $X$. For the marker attached
to the auxiliary root, this path passes through the child
endpoint of the cut edge. For the marker attached on the other
side, it passes through the parent endpoint. Thus the two
representations contain $v$ and $w$ in the first input and
$v'$ and $w'$ in the second.

In each input, remove the two marker leaves and the auxiliary
root above the child endpoint. Restore the cut edge before
applying forced contraction. This joins the two remaining
subgraphs at endpoints they already contain. Since the two
subgraphs avoid every other component representation, the
joined representation does as well.

In the first input, $\delta$ is a child of the root of
$H_\delta$, since its representation spans $\delta$ and
labels below $v$. Restoring $e$ therefore puts $H_\delta^-$
in place of $\gamma$ in $H_\gamma$ after forced contraction.
In the parallel orientation, restoring $f$ in the second
input gives the same replacement.

In the opposite orientation, restoring $f$ instead puts
$H_\gamma^-$ in place of $\delta$ in $H_\delta$. The placement
of $\gamma$ on the child side of $f$ makes it a child of the
root of $H_\gamma$. Both root-child relations hold in the
common forest $H'$. Consequently, either replacement yields
a new root with subtrees $H_\delta^-$ and $H_\gamma^-$.
Thus the joined representations give the same rooted tree in
both inputs. The rule replaces two components by one, so
the order decreases by one.

For~(ii), restricting the marked inputs to $X$ gives
$T\setminus\{e\}$ and $F\setminus\{f\}$. Thus $H'[X]$ is an
\af for $(T,F)$. Each singleton marker removed accounts for
one component, and each non-singleton marker component
survives because it contains a label in $X$. This gives
the stated order decreases.
\end{proof}

In the three-side construction, the second separation can act
on a side containing a marker from the first. Undoing the
second marker operation first gives a forest to which we can
apply the first record. Any intervening edge deletions are
already reflected in the current forest.

\begin{lemma}
\label{lem:two-level-marker-composition}
Starting from a forest pair $\mathcal I_0$, perform $t$ local
marker operations, with possible edge deletions and forced
contractions before, between, or after them. Each marker
operation acts within an independent tree--forest side and
introduces two fresh marker labels. All sides are retained
as part of the full pair. Let $\mathcal I_{\mathrm{end}}$ be
the resulting pair, and let $R_1,\ldots,R_t$ be the
reconstruction records in creation order.
Let $H_t$ be any \af for $\mathcal I_{\mathrm{end}}$.
For $j=t,t-1,\ldots,1$, restrict the current forest $H_j$
to the side modified by the $j$th marker operation, including
the two markers introduced by that operation. Apply the
reconstruction rule of Lemma~\ref{lem:maf combination for case 2.2}
to this restriction, using the marker pair and orientation
stored in $R_j$. Combine the result with the components on
the other sides to obtain $H_{j-1}$.
The resulting forest $H_0$ is an \af for $\mathcal I_0$ and
satisfies $\ord(H_0)\leq \ord(H_t)-t$.
The records occupy $O(t)$ space, and reconstruction takes
polynomial time.
\end{lemma}

\begin{proof}
If $t=0$, the terminal pair is obtained by edge deletions
and forced contraction, so its \afs are also \afs for
$\mathcal I_0$. For $t\geq1$, let $\mathcal J$ be the pair
immediately before the last marker operation. All steps
after that operation are edge deletions and forced
contractions, so $H_t$ is also an \af for the pair
immediately after the operation.

The modified side is independent of the other sides.
Apply Lemma~\ref{lem:maf combination for case 2.2} to the
restriction of $H_t$ to this side and retain the components on the
other sides. This gives an \af $H_{t-1}$ for $\mathcal J$
with $\ord(H_{t-1})\leq\ord(H_t)-1$. Only the markers of $R_t$
are removed, so all earlier markers are available. Induction
on the first $t-1$ records gives an \af $H_0$ for
$\mathcal I_0$ with
$\ord(H_0)\leq\ord(H_{t-1})-(t-1)\leq\ord(H_t)-t$.
Each record stores two marker labels and an orientation.
Each reconstruction step can be carried out by scanning the
current forest and applying forced contraction, so the total
reconstruction time is polynomial.
\end{proof}

Fix a target \af $G$ for the original pair, and suppose that
the separation steps yield a transformed target $G'$. Let $H'$
be the union of the side solutions after the selected markers
have been added back as singletons, and let $H$ be the forest
obtained by reconstructing $H'$. To show that
$\ord(H)\leq\ord(G)$, we bound $\ord(H')$ by $\ord(G')$ and
ensure that each reconstruction step removes at least as many
components as its marker operation added to the target.

If the target uses the connector, the marker operation adds
one component, and reconstruction always removes at least one.
If the target omits the connector, the operation adds two
singleton components. Reconstruction removes two components
only if both markers are still singletons when their record
is processed. A side solver may return a minimum-order forest
in which a marker belongs to a non-singleton component even
when it is a singleton in $G'$. We therefore prepare the side
inputs to ensure that selected markers are singletons when
reconstruction begins.

The algorithm tries every subset of the fresh markers as a
choice of markers to keep as singletons. For each choice,
we isolate the selected markers in both inputs of their
respective sides by deleting their incident edges whenever
present. We then remove the common singleton components,
solve each side on its remaining labels, and add the markers
back as separate components of the returned forests. The
first input to each side solver is the restriction of the
side tree to its remaining labels, which is a tree whenever
at least one label remains.

The following lemma justifies removing and restoring a
marker once it is a singleton in both inputs. It also gives
the change in the optimum order and in $\sigma_{\max}$.

\begin{lemma}
\label{lem:common-singleton-stripping}
Let $(F_1,F_2)$ be an $X$-forest pair with $|X|\geq 2$, and
suppose that $\{x\}$ is a singleton component of both inputs.
Put $X^-=X\setminus\{x\}$ and $F_i^-=F_i[X^-]$ for $i\in\{1,2\}$.
The \afs for $(F_1,F_2)$ are exactly the forests
$H^-\cup\{\{x\}\}$, where $H^-$ is an \af for $(F_1^-,F_2^-)$.
In particular, $m(F_1,F_2) = m(F_1^-,F_2^-)+1$ and $\sigma_{\max}(F_1,F_2) = \sigma_{\max}(F_1^-,F_2^-)+1$.
Consequently, $\op_3(F_1,F_2)=\op_3(F_1^-,F_2^-)$.
\end{lemma}

\begin{proof}
Every \af for $(F_1,F_2)$ has $\{x\}$ as a component because
both inputs do. Removing this component gives an \af for
$(F_1^-,F_2^-)$. Conversely, adding $\{x\}$ to an \af for
the restricted pair gives an \af for $(F_1,F_2)$. The order
increases by one under this correspondence, giving the
identity for $m$.
For $\sigma_{\max}$, apply Lemma~\ref{lem:independent-decomposition}
to the partition into $\{x\}$ and $X^-$. The restricted pair
on $\{x\}$ contributes one to $\sigma_{\max}$.
\end{proof}

The lemma lets us restore the selected markers to the side
solutions before processing the reconstruction records. To
bound the final order, the proof selects the fresh markers
that are singletons in the transformed target. Isolating
these markers preserves that target. Once restored, they
give a decrease of two for each record whose connector the
target omitted. Lemma~\ref{lem:singleton-preparation-reconstruction}
shows that any side budget sufficient for the target before
preparation is also sufficient afterward, and uses these
decreases to obtain an \af of order at most $\ord(G)$.

\subsection{Separation states and temporary budgets}
\label{subsec:separation-budgets}

Let $(T_0,F_0)$ be the tree--forest pair at the start of the
current balanced-center call, and let $k$ be its supplied budget.
For each candidate center, start the first separation with
$b=k$. Each result of this separation provides the starting
value of $b$ for the corresponding second separation. The
second separation continues to update this value. The side
solvers are called only after both separations have finished.

A separation state is $\mathcal S=(T,F;b,\mathcal R)$.
The pair $(T,F)$ is the side currently being separated,
and $\mathcal R$ is the stack of reconstruction records made
so far in the current call. Initially $(T,F)=(T_0,F_0)$ and
$\mathcal R=\varnothing$. We write $(T,F;b)$ when the stack
is understood. The initial label set is $L(T_0)$, which
includes any markers inherited from outer calls.
A marker is fresh if it is introduced by one of the two
separations at this level.

Let $c$ be the chosen nonroot internal vertex of $T_0$, and
let $e_1=(c,v_1)$ and $e_2=(c,v_2)$ be its two child connectors.
Before separating at $e_1$, store the descendant label set
$L(v_2)$ of the second connector. After the first separation,
set aside the side containing $L(v_1)$ and carry the returned
budget and reconstruction stack to the other side. The
second separation uses the edge in the remaining tree whose
descendant label set is $L(v_2)$. For example, if the first
separation cuts $e_1$ without adding markers, forced
contraction suppresses $c$ and joins its parent directly to
$v_2$. Lemma~\ref{lem:current-connector-representative}
identifies the required edge whether or not the first
separation introduces markers.

The budget changes according to the operation performed:
\begin{itemize}
  \item Each essential edge deletion in the second input
        by the two-color process of Section~\ref{sec:difficult}
        is \emph{budget-charged}. This includes deletions
        made by Reduction Rule~0 in that process. The
        deletion increases the second-input order by one
        and decreases $b$ by one.
  \item Each marker operation increases the second-input
        order by one and adds one record to $\mathcal R$.
        Increase $b$ by one and exclude the edge deleted
        by this operation from the count of budget-charged
        deletions.
  \item An unmarked connector deletion in the first input
        leaves $b$ and $\mathcal R$ unchanged.
\end{itemize}
On a branch of the search, let $i$ count the charged
deletions and let $t=|\mathcal R|$. Both counts include
operations from the first and second separations. Thus
$b=k+t-i$.

For the correctness proof, fix an \af $G$ for $(T_0,F_0)$.
A sequence is \emph{target-compatible} if each unmarked
deletion preserves the current target and each marker
operation transforms it by the corresponding
target-preservation lemma. If the current target uses the
connector, the marker operation splits one target component
into two. If it omits the connector, the operation adds
two singleton components. The second-input order increases
by one in either case, so the difference between the target
order and the second-input order can increase by at most one.
This is the reason for the update $b\leftarrow b+1$.
The algorithm applies the same budget rules to every branch.

At any point in this sequence, write $F'$ for the union of
the current second-input forests, including the forest on any
side already set aside. Write $G'$ for the transformed target
on all these sides. Target compatibility ensures that $G'$
is an \af for the current inputs on those sides.
In particular, the order of $F'$ counts
all second-input components, even when the state $(T,F)$
contains only the side being separated.

\begin{lemma}
\label{lem:temporary-budget-invariant}
Suppose that $G$ is an \af for $(T_0,F_0)$ with
$\ord(G)\leq\ord(F_0)+k$ and that the separation steps so
far are target-compatible. Let $q$ count the records
created at steps where the current target omitted the
connector. Before isolating any markers for the side calls,
we have
\[
  b=k+t-i,\qquad
  \ord(F')=\ord(F_0)+i+t,\qquad
  \ord(G')=\ord(G)+t+q.
\]
In particular, $0\leq\ord(G')-\ord(F')\leq b$.
\end{lemma}

\begin{proof}
The budget updates give the first identity. Each of the
$i$ charged deletions and $t$ marker operations increases
the order of the second input by one. An unmarked deletion
preserves the target order. The $q$ marker operations whose
connectors are omitted add two components each, and the
remaining $t-q$ operations add one component each. This
gives the other two identities. Since $q\leq t$ and the
initial target satisfies $\ord(G)-\ord(F_0)\leq k$,
\[
  \ord(G')-\ord(F')
  =\ord(G)-\ord(F_0)+q-i
  \leq k+t-i=b.
\]
The difference is nonnegative because $G'$ is a subforest
of $F'$. Hence the budget cannot become negative along
such a sequence.
\end{proof}

After the first separation, $\ord(G')-\ord(F')$ is the
sum of the two side differences between target order and
second-input order. Both are nonnegative, so each is at
most the returned value of $b$. In particular, this value
suffices for the target on the side that undergoes the
second separation.

Once the separations are complete, let $(T_\ell,F_\ell)$
be the pair on side $\ell$, with label set $Y_\ell$.
For each subset $S$ of the fresh markers, the algorithm
tests the choice of keeping all markers in $S$ as singleton
components. On side $\ell$, put $S_\ell=S\cap Y_\ell$.
Isolate every marker in $S_\ell$ in both inputs by deleting
its incident edge whenever present, and remove the resulting
singleton components. The pair passed to the side solver is
$(\widetilde T_\ell,\widetilde F_\ell)
  =
  (T_\ell[Y_\ell\setminus S_\ell],
        F_\ell[Y_\ell\setminus S_\ell])$.
We call this the \emph{prepared side pair}. Each call uses
the additional-cut budget $\kappa$, so it seeks an \maf of
order at most $\ord(\widetilde F_\ell)+\kappa$ and reports
failure if no such forest exists. The balanced-center
algorithm sets $\kappa=\min\{\lfloor k/2\rfloor+1,b\}$ using
the final value of $b$ for that branch. If all side calls
succeed, add back the removed markers as singleton components
of their respective solutions.

In the proof we choose $S$ to contain exactly the fresh
markers that are singletons in $G'$. This preserves the
target when the markers are isolated. It also ensures that
a record made when the target omitted its connector will
remove two singleton components during reconstruction.

\begin{lemma}
\label{lem:singleton-preparation-reconstruction}
Under the assumptions of Lemma~\ref{lem:temporary-budget-invariant},
suppose the separation has produced $s$ independent
tree--forest sides $(T_\ell,F_\ell)$ on label sets $Y_\ell$,
for $\ell=1,\ldots,s$. Suppose each $Y_\ell$ contains at least
one label of $T_0$. Let $S$ consist of the fresh markers that
are singleton components of $G'$, and let $S_\ell=S\cap Y_\ell$.
Write $G_\ell=G'[Y_\ell]$ and
$\widetilde G_\ell=G_\ell[Y_\ell\setminus S_\ell]$.
Use the side inputs $(\widetilde T_\ell,\widetilde F_\ell)$
obtained by the procedure just described.
\begin{enumerate}[(i)]
  \item For every $\ell$, $\widetilde T_\ell$ is a nonempty
        tree and $\widetilde G_\ell$ is an \af for
        $(\widetilde T_\ell,\widetilde F_\ell)$. Moreover,
        \[
          \ord(\widetilde G_\ell)-\ord(\widetilde F_\ell)
          \leq \ord(G_\ell)-\ord(F_\ell).
        \]
        Thus a budget sufficient for $G_\ell$ on the side
        before removing the markers also suffices for
        $\widetilde G_\ell$ on the pair given to the solver.
        Isolating and removing the selected markers leaves
        $b$ and $\mathcal R$ unchanged.

  \item Let $\widetilde H_\ell$ be an \maf for
        $(\widetilde T_\ell,\widetilde F_\ell)$, and set
        $H'=\bigcup_{\ell=1}^s
          \bigl(\widetilde H_\ell\cup\{\{x\}:x\in S_\ell\}\bigr)$.
        Then $\ord(H')\leq\ord(G')$. Processing the records
        in reverse order gives an \af $H$ for $(T_0,F_0)$
        satisfying
        \[
          \ord(H)\leq\ord(H')-(t+q)
          \leq\ord(G)\leq\ord(F_0)+k.
        \]
\end{enumerate}
\end{lemma}

\begin{proof}
For~(i), each marker in $S_\ell$ is already a singleton
component of $G_\ell$. Isolating these markers in the two
inputs therefore preserves $G_\ell$. Removing them gives
the target $\widetilde G_\ell$ of order
$\ord(G_\ell)-|S_\ell|$. In the second input, isolating the
markers can only increase the order or leave it unchanged,
and removing them decreases the order by $|S_\ell|$.
Hence $\ord(\widetilde F_\ell)\geq\ord(F_\ell)-|S_\ell|$.
Subtracting gives the stated inequality.
The restriction of $T_\ell$ is a nonempty tree because
$S_\ell$ contains only fresh markers and at least one label
of $T_0$ remains on side $\ell$. With a tree as the first
input, the order differences in~(i) are exactly the
additional-cut requirements for the respective targets.
The marker removals take place after separation and do not
contribute to either count in $b=k+t-i$.

For~(ii), $\ord(\widetilde H_\ell)\leq\ord(\widetilde G_\ell)$.
By Lemma~\ref{lem:common-singleton-stripping}, adding back
the selected markers as singletons gives an \af for the
side with those markers isolated, and hence for
$(T_\ell,F_\ell)$. Its order is at most
$\ord(\widetilde G_\ell)+|S_\ell|=\ord(G_\ell)$.
The sides are independent, so their union $H'$ is an \af
for the pair after both separations and satisfies
$\ord(H')\leq\sum_{\ell=1}^s\ord(G_\ell)=\ord(G')$.

Each of the $q$ records counted in
Lemma~\ref{lem:temporary-budget-invariant} added two
singleton components to the current target. Later
target-compatible steps leave those components intact,
so both markers belong to $S$ and are singletons in $H'$.
A reconstruction step for another record acts only on
components containing its own markers. The two singletons
therefore remain until their own record is processed.
By Lemma~\ref{lem:maf combination for case 2.2}, each of
these $q$ records removes two components and each of the
other $t-q$ records removes at least one.
Lemma~\ref{lem:two-level-marker-composition} guarantees
that the resulting $H$ is an \af for $(T_0,F_0)$. Thus
\[
  \ord(H)\leq\ord(H')-(t+q)
  \leq\ord(G')-(t+q)=\ord(G)\leq\ord(F_0)+k,
\]
where the equality follows from
Lemma~\ref{lem:temporary-budget-invariant}.
\end{proof}

Each of the two separations creates at most one record
on a branch. Thus for each result of the two separations,
the algorithm tries at most $2^4=16$ subsets of its fresh
markers. These include the set $S$ in the proof.
Lemma~\ref{lem:balanced-center-interface} establishes
that, for a suitable center and separation, the side
budget $\kappa$ suffices for this choice.

\section{Difficult Binary Separation}
\label{sec:difficult}

The easy constructions apply when at most one edge deletion
from the second input separates the prescribed sides.
We now consider the remaining case.

Let $T$ and $F$ be the tree and forest for the current
separation, with temporary budget $b\geq0$.
We write $(T,F;b)$ for this input, omitting the
reconstruction stack as in
Section~\ref{subsec:separation-budgets}.
The two inputs have the same label set $X$, including
any marker labels already present.
Let $e=(w,v)$ be an edge of $T$, where $w$ is the parent
of $v$, and let $X_v=L(v)$ and $X_w=X\setminus X_v$.
The separation is \emph{difficult} if $F$ does not admit
$(X_v,X_w)$ and no single edge deletion makes it admit
this bipartition.

For the correctness proof, fix an \af $G$ satisfying
$\ord(G)\leq\ord(F)+b$.
If $e\notin E_T(G)$, deleting $e$ preserves $G$.
If $e\in E_T(G)$, one target component crosses the
bipartition, and we use the local marker operation to
retain the information needed to reconnect its two parts.
These give the unmarked and marked constructions,
respectively.

In either construction, further cuts in the second input
may be needed.
The two-color process below makes these cuts and preserves
a feasible target on at least one branch.
Subsection~\ref{subsec:difficult-interface} gives the complete
procedure combining the two constructions.

\subsection{The two-color separation process}

We first define the input of this process independently
of the initial construction that produced it. Let $U$ be the full label
set at the start of the process. Let $F_1$ be a rooted
binary $U$-forest with exactly two components, whose
nonempty label sets are $Y$ and $Z$, so that
$U=Y\mathbin{\dot\cup}Z$. Let $F_2$ be a rooted binary
$U$-forest on the same label set. The first forest $F_1$
is kept fixed throughout the process. We delete edges
only from $F_2$, with the aim of making every component
have all its labels in $Y$ or all in $Z$.

For the unmarked construction, these inputs are
$F_1=T\setminus\{e\}$, $F_2=F$, $U=X$, $Y=X_v$, $Z=X_w$.
For a marked construction, let $(F_T,F_F)$ be the pair obtained
from the chosen edge and orientation. Then
$F_1=F_T$, $F_2=F_F$, $U = X\cup\{\delta,\gamma\}$, $Y=X_v\cup\{\delta\}$, $Z=X_w\cup\{\gamma\}$.
The two marker orientations change the placement of the
markers in $F_F$, but not in $F_T$. Thus $Y$ and $Z$ are
always the label sets of the two components of $F_1$.
In particular, fresh markers are included in the color
classes. In the rest of this subsection, $b$ denotes the
nonnegative integer budget supplied to the process,
after any marker credit has been added.

To identify useful cuts, we consider the subtrees of $F_2$
rooted at its vertices.
We refer to these as \emph{local subtrees} in this subsection.
A local subtree is \emph{monochromatic} if all its labels
belong to $Y$ or all belong to $Z$.
It is \emph{maximal monochromatic} if it is not properly
contained in another monochromatic local subtree.
These maximal subtrees are disjoint and cover the labels.
Construct $P(F_2)$ by replacing each maximal monochromatic
subtree with a leaf. For a leaf $l$ of $P(F_2)$, let $B(l)$
be the label set of the subtree it represents. The leaf $l$
has color~$0$ if $B(l)\subseteq Y$ and color~$1$ if
$B(l)\subseteq Z$.

The forest $P(F_2)$ is used to choose cuts. The actual
input $F_2$ retains all labels and all edges inside the
contracted subtrees. Every edge of $P(F_2)$ corresponds
to one edge of the current $F_2$ outside those subtrees.
When a rule specifies an edge of $P(F_2)$, the algorithm
deletes this corresponding edge from $F_2$. After deleting
the edges selected by a rule, it applies forced contraction
to $F_2$ and rebuilds $P(F_2)$. In particular, monochromatic
subtrees are recomputed from the current forest after
every recursive step.

\begin{lemma}
\label{lem:PF-bichromatic-siblings}
No two sibling leaves of $P(F_2)$ have the same color.
\end{lemma}

\begin{proof}
If two sibling leaves had the same color, the local subtree
rooted at their parent would be monochromatic. It would
properly contain both of the maximal monochromatic subtrees
represented by those leaves, contradicting maximality.
\end{proof}

Before applying a branching rule, we apply Reduction Rule~0
to $F_2$ with respect to the fixed forest $F_1$. Each
reduction is performed as a separate budget-charged step:
one essential edge is deleted and the budget decreases
by one. This reduction stage has the following useful
consequence for the colored forest.

\begin{lemma}
\label{lem:reduced-bichromatic-components}
Suppose that Reduction Rule~0 is not applicable to $F_2$
with respect to $F_1$. Every component of $P(F_2)$ containing
both colors has at least two leaves of color~$0$ and at least
two leaves of color~$1$.
\end{lemma}

\begin{proof}
Let $C$ be a component of $F_2$ whose corresponding
component in $P(F_2)$ contains both colors. Suppose that
there is exactly one leaf $l$ of color~$0$. Its represented
subtree contains every label of $C$ belonging to $Y$, so
$B(l)=L(C)\cap Y$. This subtree is a proper subtree of
$C$, because $C$ also contains labels from $Z$.
Let $f$ be the edge entering its root. Deleting $f$
separates the nonempty label sets $B(l)$ and
$L(C)\setminus B(l)$, so $f$ is essential. Since $Y$ is
the label set of one component of $F_1$, the equality
$B(l)=L(C)\cap Y$ makes Reduction Rule~0 applicable to
$f$ in $F_2$ with respect to $F_1$, a contradiction.
The argument with $Y$ and $Z$ exchanged excludes a unique
leaf of color~$1$.
\end{proof}

The branching proofs use two consequences of the \af definition. 
First, every component of an \af $H$
for $(F_1,F_2)$ has all its labels in $Y$ or all in $Z$,
because it is a subforest of $F_1$. Second, its distinct
component representations in $F_2$ are vertex-disjoint.
Consequently, if two sibling leaves of $P(F_2)$ have
different colors, their two incoming edges cannot both
occur in the representation of $H$. These edges share a parent. 
If two components of $H$
used these edges, their representations would intersect
at that parent. If one component used both edges, it would
contain labels of both colors.

We will also use minimality of the representations. If
the edge entering a represented subtree occurs in $H$,
the component using that edge must contain a label inside
the subtree and a label outside it. Otherwise the edge
would not be part of its minimal representation. Here
and below, occurrence of an edge in $H$ refers to its
representation in the current second forest $F_2$.

\medskip
\noindent\textbf{Branching Rule~1.}
Suppose that $P(F_2)$ has a local subtree with exactly
three leaves $l_1,l_2,l_3$, where $l_1,l_2$ are siblings
and $l_1,l_3$ have the same color. For $i\in\{1,2,3\}$,
let $f_i$ be the edge of $F_2$ corresponding to the edge
entering $l_i$. Create the following two branches:
\begin{enumerate}[(B1)]
  \item Delete $f_1$ and $f_3$, using two units of budget.
  \item Delete $f_2$, using one unit of budget.
\end{enumerate}
Discard a branch if its cost exceeds the remaining budget.

\begin{lemma}
\label{lem:Branching Rule-1}
Let $H$ be an \af for the current pair $(F_1,F_2)$.
At least one of the cut sets of Branching Rule~1 is
disjoint from the representation of $H$ in $F_2$.
The branching vector is $(2,1)$.
\end{lemma}

\begin{proof}
Let $B_i=B(l_i)$. Let $a$ be the parent of $l_1,l_2$
in $P(F_2)$, and let $z$ be the root of the three-leaf
local subtree. Thus $z$ has children $a$ and $l_3$.
The vertices $a,z$ are also vertices of $F_2$, outside
the contracted monochromatic subtrees.

If $f_2$ does not occur in $H$, deleting it preserves
$H$, giving Branch~(B2). Suppose instead that $f_2$ occurs.
The sibling-edge observation above gives $f_1\notin E_{F_2}(H)$.
Let $D_2$ be the component of $H$ whose representation
uses $f_2$. By minimality, $D_2$ has a label in $B_2$
and a label outside $B_2$. It has no label in $B_1$,
since $B_1$ and $B_2$ have different colors. Its
representation must therefore extend from the $B_2$
subtree through $a$ to $z$.

If $f_3$ also occurred in $H$, the component $D_3$ using
it would contain $z$, the parent endpoint of $f_3$.
The components $D_2,D_3$ cannot be distinct, since their
representations would intersect at $z$. They cannot be
the same component either, since $B_2$ and $B_3$ have
different colors. Hence $f_3\notin E_{F_2}(H)$, and deleting
$f_1,f_3$ preserves $H$, giving Branch~(B1).

The two-edge cut detaches the nonempty subtrees with
label sets $B_1,B_3$ and leaves a nonempty component
containing $B_2$. Its deletion increases the forest order
by exactly two, so the cut set is an ee-set. Deleting
$f_2$ alone also separates two nonempty label sets and
increases the order by one. These are the stated budget
costs.
\end{proof}

\medskip
\noindent\textbf{Branching Rule~2.}
Suppose that $P(F_2)$ has a local subtree with exactly
four leaves $l_1,l_2,l_3,l_4$, where $l_1,l_2$ are siblings,
$l_3,l_4$ are siblings, $l_1,l_3$ have the same color,
and $l_2,l_4$ have the same color. For each $i$, let $f_i$
be the edge of $F_2$ entering the root of the subtree
represented by $l_i$.
Create the following two
branches.
\begin{enumerate}[(B1)]
  \item Delete $f_1$ and $f_3$, using two units of budget.
  \item Delete $f_2$ and $f_4$, using two units of budget.
\end{enumerate}
Discard a branch if its cost exceeds the remaining budget.

\begin{lemma}
\label{lem:Branching Rule-2}
Let $H$ be an \af for the current pair $(F_1,F_2)$.
At least one of the cut sets of Branching Rule~2 is
disjoint from the representation of $H$ in $F_2$.
The branching vector is $(2,2)$.
\end{lemma}

\begin{proof}
Let $B_i=B(l_i)$, and let $z$ be the root of the
four-leaf local subtree. The two children of $z$ are
the parents of the two sibling pairs. The incoming
edges of either sibling pair cannot both occur in $H$.

Suppose that neither proposed cut set is disjoint from
$H$. Then $H$ uses an edge from $\{f_1,f_3\}$ and an
edge from $\{f_2,f_4\}$. These two edges cannot belong
to the same sibling pair. Up to exchanging the two pairs,
we may therefore assume that $f_1,f_4$ occur in $H$,
whereas $f_2,f_3$ do not.

Let $D_1,D_4$ be the components using $f_1,f_4$,
respectively. Minimality forces the representation of
$D_1$ to extend outside the $B_1$ subtree. It cannot
extend into $B_2$, which has the opposite color, so
it must pass through $z$. Likewise, the representation
of $D_4$ must pass through $z$, because it cannot extend
into the oppositely colored $B_3$ subtree. If $D_1,D_4$
are distinct, their representations intersect at $z$.
If they coincide, that component contains labels from
the differently colored sets $B_1$ and $B_4$.
Both possibilities contradict the \af
conditions. One of the proposed cut sets therefore
preserves $H$.

Each cut set detaches two nonempty represented subtrees
and leaves the other two in a nonempty remaining component.
It is an ee-set of size two and increases the forest
order by exactly two.
\end{proof}

The two branching lemmas identify cut sets that preserve
a target. The procedure explores only those cut sets whose
cost fits within the remaining budget. For example, with
one unit left, it can explore only the one-edge branch of
Branching Rule~1. Lemma~\ref{lem:2LS-Process-1} proves that
a feasible target has a preserving branch within budget.

To complete the separation process, we must also show
that these local rules are available whenever a
bichromatic component remains after Reduction Rule~0.
The following structural observation supplies this step.

\begin{lemma}
\label{lem:two-color-terminal}
Let $C$ be a rooted binary tree whose leaves each have color~$0$ or~$1$. 
If $C$ contains at least two leaves of each
color and no two sibling leaves have the same color, then
Branching Rule~1 or Branching Rule~2 applies to a local
subtree of $C$.
\end{lemma}

\begin{proof}
Choose a deepest vertex whose local subtree has at least
three leaves. Neither child subtree can have three or
more leaves, by the choice of this vertex. Thus the two
child subtrees have sizes $(1,2)$ or $(2,2)$.
In the first case, the size-two subtree is a sibling
pair with different colors. The remaining leaf shares
its color with one of this pair, giving Branching Rule~1.
In the second case, both child subtrees are sibling
pairs with different colors. Number the leaves within
the pairs so that $l_1,l_3$ have the same color, giving
Branching Rule~2.
\end{proof}

The procedure in Algorithm~\ref{alg:2LS-Process-1} applies
Reduction Rule~0 before branching. If several branching
configurations are available, it chooses one using a
fixed tie-breaking rule and creates only the branches
of that configuration. Each deleted edge consumes one
unit of budget. The procedure skips cuts that exceed the
remaining budget and returns only states in which every
component of the second forest has labels of a single
color.

\begin{algorithm}[htbp]
\caption{\algname{2LS-Process-1}$(F_1,F_2;b)$}
\label{alg:2LS-Process-1}
\small

\textbf{Input:}
Two $U$-forests $F_1,F_2$ and a nonnegative integer budget $b$.
The forest $F_1$ has exactly two components with nonempty
label sets $Y$ and $Z$.\\
\textbf{Output:}
A list $\mathcal C$ of terminal states $(F_1,F_2';b')$.
Every returned state satisfies the following conditions:
\begin{itemize}
  \item $F_2'$ is obtained from $F_2$ by edge deletions
        and forced contraction;
  \item every component of $F_2'$ has all its labels
        in $Y$ or all in $Z$;
  \item the remaining budget satisfies $0\leq b'\leq b$ and
        $b-b'=\ord(F_2')-\ord(F_2)$.
\end{itemize}

\medskip
\algrenewcommand{\algorithmicfunction}{}
\algrenewcommand{\algorithmicprocedure}{}
\algtext*{EndFunction}
\algtext*{EndProcedure}
\begin{algorithmic}[1]
  \Function{2LS-Process-1}{$F_1,F_2;b$}
    \State $\mathcal C\gets[\,]$;
    \State \Call{Explore}{$F_1,F_2;b,\mathcal C$};
    \State \Return $\mathcal C$.
  \EndFunction

  \Statex
  \Statex \emph{// All calls to \textsc{Explore} in this search share
the same list $\mathcal C$.}
  \Procedure{Explore}{$F_1,F_2;b,\mathcal C$}
    \If{Reduction Rule~0 applies to $F_2$ with respect to $F_1$}
      \If{$b\geq1$}
        \State choose one applicable edge $f$ using a fixed rule;
        \State \Call{Explore}
          {$F_1,F_2\setminus\{f\};b-1,\mathcal C$};
      \EndIf
      \State \Return.
    \EndIf

    \State construct $P(F_2)$;
    \If{no branching rule applies to $P(F_2)$}
      \State append $(F_1,F_2;b)$ to $\mathcal C$;
      \State \Return.
    \EndIf

    \State choose one applicable configuration using a fixed rule.
    \For{each of the two cut sets $D$ prescribed by this configuration}
      \If{$|D|\leq b$}
        \State \Call{Explore}
          {$F_1,F_2\setminus D;b-|D|,\mathcal C$};
      \EndIf
    \EndFor
  \EndProcedure
\end{algorithmic}
\end{algorithm}

Every recursive step deletes at least one essential edge
from $F_2$, so its order strictly increases. Since the
order is at most $|U|$, every search path is finite.
We next verify that a returned terminal state has achieved
the required separation.

\begin{lemma}
\label{lem:property of 2LS-Process-1}
For every terminal state $(F_1,F_2';b')$ returned by
\textsc{2LS-Process-1}, every component of $F_2'$ has
all its labels in $Y$ or all in $Z$. Equivalently, the component-label partition of $F_2'$ refines
that of $F_1$.
\end{lemma}

\begin{proof}
Suppose that a component of $P(F_2')$ contains both colors.
Reduction Rule~0 is inapplicable at a terminal state,
so Lemma~\ref{lem:reduced-bichromatic-components}
gives at least two leaves of each color in this component.
Lemma~\ref{lem:PF-bichromatic-siblings} excludes sibling
leaves of the same color. Lemma~\ref{lem:two-color-terminal}
then supplies a branching configuration, contradicting
terminality. Every component is therefore monochromatic,
which is exactly the required statement about its labels.
\end{proof}

The next lemma shows that the budget tests do not remove
every branch preserving a feasible target. This requires
both the preservation arguments above and the fact that
the stated cut costs equal the increases in forest order.

\begin{lemma}
\label{lem:2LS-Process-1}
Let $H$ be an \af for $(F_1,F_2)$ satisfying
$\ord(H)\leq\ord(F_2)+b$. Some terminal state
$(F_1,F_2';b')$ returned by \textsc{2LS-Process-1}
preserves $H$ and satisfies
$\ord(H)\leq\ord(F_2')+b'$.
\end{lemma}

\begin{proof}
Follow a branch that preserves $H$. A Reduction Rule~0
step preserves every \af by
Lemma~\ref{jianer-1}. If a branching rule is applied,
Lemma~\ref{lem:Branching Rule-1} or
Lemma~\ref{lem:Branching Rule-2} supplies a cut set
disjoint from the representation of $H$.

Let $\ell$ be the cost of this preserving step, and write
$\widehat F_2$ for its resulting second forest.
The deleted edges form an ee-set, so
$\ord(\widehat F_2)=\ord(F_2)+\ell$. Since $H$ remains
an \af, its order is at least
$\ord(\widehat F_2)$. Thus
\[
  \ord(F_2)+\ell
  =\ord(\widehat F_2)
  \leq\ord(H)
  \leq\ord(F_2)+b,
\]
which implies $\ell\leq b$. This preserving step is
therefore not discarded by the budget test. Moreover,
$\ord(\widehat F_2)+(b-\ell)=\ord(F_2)+b$,
so the feasibility inequality holds at the child call.
Apply the same argument at each subsequent call.
The search path is finite, and it cannot end in a
budget rejection while preserving $H$. It therefore
reaches a returned terminal state with the claimed
properties.
\end{proof}

We finally bound the branching cost. The weighted form
of the bound records the number of deletions used by
each terminal state. This is the information needed when
separation is later combined with the side solvers.

\begin{lemma}
\label{lem:2LS-Process-1 branching complexity}
Let $\varphi=(1+\sqrt5)/2$. For a call with initial
budget $b$, let $\mathcal C_i$ be the collection of
returned terminal states reached after exactly $i$
budget-charged deletions, counting calls reached by
different search paths separately. Then $\sum_i |\mathcal C_i|\varphi^{-i} \leq 1$.
In particular, the process has branching factor at most
$\varphi$ and runs in $O^*(\varphi^b)$ time.
\end{lemma}

\begin{proof}
Give a search node reached after $i$ deletions weight
$\varphi^{-i}$. For Branching Rule~1, the two child
weights sum to the parent weight because
$\varphi^{-1}+\varphi^{-2}=1$. For Branching Rule~2,
their sum is smaller because $2\varphi^{-2}<1$.
A deterministic Reduction Rule~0 step has one child
with weight $\varphi^{-1}$ times the parent weight.
Discarding a child whose cost exceeds the remaining
budget cannot increase the sum. Hence the total weight
of all leaves, including rejected leaves, is at most
the root weight, which is one. The returned terminal
states form a subcollection of these leaves, proving
the displayed inequality.

Every generated call has nonnegative budget, so a leaf
is reached after at most $b$ deletions and has weight
at least $\varphi^{-b}$. There are therefore at most
$\varphi^b$ leaves. Every search path has length at
most $|U|$, and the work at each node is polynomial
in $|U|$. The resulting running time is
$O^*(\varphi^b)$.
\end{proof}

The caller now has two independent sides, $(F_1[Y],F_2'[Y])$
and $(F_1[Z],F_2'[Z])$.
The first input of each side is a tree. The caller can
retain a side for a later separation or solve it when
all required separations are complete. The present
procedure supplies the separated inputs and the remaining
budget; side solving and reconstruction are performed
by the outer construction.

Since each component of $F_2'$ lies within one of the
two component label sets of $F_1$, the component-pair
partition consists exactly of the component label sets
of $F_2'$. Thus $\sigma_{\max}(F_1,F_2')=\ord(F_2')$.
For a preserved target $H$, the remaining feasibility
inequality can therefore be written as
$\ord(H)-\sigma_{\max}(F_1,F_2')\leq b'$.
This explains why the remaining additional-cut budget
is also a refined-gap budget at a terminal state.

\subsection{The case where the target contains the connector}
\label{subsec:target-contains-e}

Suppose that the fixed target $G$ uses $e$, and let $C_e$
be the component whose representation in $T$ contains $e$.
We seek a corresponding cut in $F$ that allows $C_e$
to be replaced by two compatible marked components.

Let $(F_T,F_F)$ be the pair obtained by applying the
local marker operation to $e$ and a chosen edge of $F$.
The two components of $F_T$ have label sets
$Y=X_v\cup\{\delta\}$ and $Z=X_w\cup\{\gamma\}$.
We show that there is an edge of $F$ for which the
parallel orientation preserves a transformed target.
The two-color process then separates any components of
$F_F$ that still have labels from both sides.

\begin{lemma}
\label{lem:marked-connector-guess}
Let $G$ be an \af for $(T,F)$ whose representation in
$T$ contains $e$. There is an edge $f\in E(F)$ such that
the local marker operation on $e$ and $f$ in the parallel
orientation gives a pair $(F_T,F_F)$ with an \af
$\widehat G$ of order $\ord(G)+1$.
Every component of $\widehat G$ has all its labels in
$Y$ or all in $Z$, and both fresh markers belong to
non-singleton components of $\widehat G$.
\end{lemma}

\begin{proof}
First, $C_e$ is the only component of $G$ containing
labels from both $X_v$ and $X_w$. The representation in
$T$ of any component meeting both sides must contain
$e$. Two such components would therefore have intersecting
representations, which an \af does not allow.
Let $A_v=L(C_e)\cap X_v$ and $A_w=L(C_e)\cap X_w$.
Both sets are nonempty because $e$ belongs to the minimal
representation of $C_e$.

Let $R_T(C_e)$ and $R_F(C_e)$ be the minimal
representations of $C_e$ in the two inputs. Suppressing
their subdivision vertices gives the same rooted tree
$C_e$. In $R_T(C_e)$, the edge $e$ lies on the path
representing the directed edge $\eta$ of $C_e$ with
descendant label set $A_v$. Let $P_\eta$ be the path
representing the same edge in $R_F(C_e)$. The internal
vertices of $P_\eta$ have only one child in $R_F(C_e)$.
Thus, for each edge of $P_\eta$, the labels of $C_e$
below that edge in $F$ are exactly $A_v$. Fix one such
edge $f$.

Cutting $e$ and $f$ puts $A_v$ on the child side and
$A_w$ on the parent side in both representations.
The parallel orientation therefore assigns $\delta$
to the part containing $A_v$ and $\gamma$ to the part
containing $A_w$ in both inputs. Define $\widehat G$
by applying the local marker operation to $\eta$ in
$C_e$ with these marker placements. Replace $C_e$
by the resulting two trees, with label sets
$A_v\cup\{\delta\}$ and $A_w\cup\{\gamma\}$, and retain
the other components of $G$.

We verify that these two trees are represented in both
marked inputs. In $T$, the marker operation cuts the
path representing $\eta$ at $e$; in $F$, it cuts the
corresponding path at $f$. In each representation,
$\delta$ is placed at the cut end on the descendant
side and $\gamma$ at the cut end on the parent side.
After the unused edges are omitted, forced contraction
suppresses the remaining subdivision vertices on these
paths. The marker attachments therefore reduce to those
specified at $\eta$ in $C_e$. The resulting two rooted
trees are the same in both inputs.

There may be subtrees containing labels of other target
components attached along $P_\eta$ in the full forest
$F$. We retain the representations of all those components.
They are vertex-disjoint from $R_F(C_e)$ and hence use
no vertex of $P_\eta$. The new marker attachments use
only vertices of $R_F(C_e)$ and newly inserted vertices,
so they do not intersect these retained representations.
We omit only edges and vertices not needed by the
component representations; no labels or components of
$G$ are discarded. The same disjointness argument applies
to the representations in $T$. Thus $\widehat G$ is
a common subforest of $F_T$ and $F_F$.

Replacing $C_e$ by two components increases the order
by one. Each of the two new components contains a
nonempty set of labels from $C_e$ as well as its marker,
so both marker components are non-singletons. These
components lie on the prescribed sides $Y,Z$. Every other component
already lies in $X_v$ or $X_w$ and is unchanged. This
proves all the claims.
\end{proof}

\subsection{The difficult-separation procedure}
\label{subsec:difficult-interface}

The preceding lemmas distinguish whether the target uses
the connector $e$. Since the algorithm does not know the
target, it tries both constructions.

The input is a tree--forest pair $(T,F)$, a temporary
budget $b$, and a connector $e$ inducing the bipartition
$(X_v,X_w)$.
The procedure returns separated candidate states, each
with its remaining budget and reconstruction records.
Every choice below starts from a separate copy of the
incoming state, including its existing records.

\begin{enumerate}[(i)]
  \item \label{item:difficult-unmarked}\emph{Separation without markers.}
        Delete $e$ from $T$ and call
        \textsc{2LS-Process-1} on
        $(T\setminus\{e\},F;b)$.
        The prescribed sides are $X_v$ and $X_w$.
        This construction introduces no marker and
        adds no reconstruction record.

  \item \label{item:difficult-marked}\emph{Separation with markers.}
        For every edge $f\in E(F)$, apply the local marker
        operation to $e$ and $f$ in the parallel orientation,
        obtaining $(F_T,F_F)$. Append a reconstruction
        record for the fresh marker pair and the parallel
        orientation, and call
        \textsc{2LS-Process-1} on $(F_T,F_F;b+1)$.
        The prescribed sides are
        $Y=X_v\cup\{\delta\}$ and $Z=X_w\cup\{\gamma\}$.
\end{enumerate}

Retain all terminal states returned by these calls,
together with their remaining budgets and records.
No side solver is called at this stage.
The outer algorithm first completes both binary
separations, then solves the resulting sides and
applies the reconstruction rules.

Fix an \af $G$ for $(T,F)$ with
$\ord(G)-\ord(F)\leq b$.
If its representation in $T$ omits $e$, deleting $e$ as in
item~(\ref{item:difficult-unmarked}) preserves $G$.
The second forest remains $F$, so $G$ is feasible for the
two-color call with budget $b$.
By Lemma~\ref{lem:2LS-Process-1}, this call returns a state
that preserves $G$.

If the representation of $G$ in $T$ contains $e$,
Lemma~\ref{lem:marked-connector-guess} gives an edge $f$
for which the marker operation in item~(\ref{item:difficult-marked})
produces a pair $(F_T,F_F)$ admitting an \af $\widehat G$
of order $\ord(G)+1$. The marked forest $F_F$ also has
one more component than $F$, so
$\ord(\widehat G)-\ord(F_F) =\ord(G)-\ord(F)\leq b$.
The budget $b+1$ used for the marked call therefore
suffices for $\widehat G$. By
Lemma~\ref{lem:2LS-Process-1}, this call returns a state
that preserves $\widehat G$.

Lemma~\ref{lem:deferred-binary-interface} collects the
separation guarantees, target-preservation properties,
budget updates, and weighted bounds for this procedure
and the easy-separation constructions.

\section{The Balanced-Center Algorithm}
\label{sec:balanced-center-algorithm}

For a fixed target, a balanced center gives three label
sides whose restricted target gaps satisfy the bounds of
Section~\ref{sec:balanced-centers}. These restrictions
are taken before fresh markers are added. To use them
in the algorithm, we separate the inputs along the same
sides, solve the resulting instances, and reconstruct an
\af of order no larger than the target.
Since the target and its critical edge set are unknown,
the algorithm enumerates all nonroot internal vertices
as candidate centers.

At a candidate center, the algorithm performs two binary
separations. It stores both child connectors, separates
along the first, and sets the separated side aside. It
then separates the remaining side along the second
connector. Only after both stages have finished does it
solve the three side instances and reconstruct their
solutions. The two stages use the same budget and count
of charged deletions, and create at most two new
reconstruction records. The fresh markers also affect
the side gaps; this effect must be included when bounding
the parameters passed to the side solvers.

The first separation may change the edge representing the second
connector. The following lemma identifies it using its descendant
labels, as described in Subsection~\ref{subsec:separation-budgets}.

\begin{lemma}
\label{lem:current-connector-representative}
Let $c$ be a nonroot internal vertex of the current tree $T$,
with children $v_1,v_2$ and parent $p$. Put $X_2=L(v_2)$.
Perform the first separation at $e_1=(c,v_1)$ and retain its
root-side tree $R$. Then $R$ has a unique edge with descendant
label set $X_2$. This is the current representative of
$e_2=(c,v_2)$ and can be found in polynomial time.
\end{lemma}

\begin{proof}
The first separation changes the first input only at $e_1$.
Without markers, it cuts $e_1$ and suppresses $c$; the edge
$(p,v_2)$ then has descendant set $X_2$. With markers, it
replaces the child side at $c$ by the fresh leaf $\gamma$.
The vertex $c$ retains two children, $v_2$ and $\gamma$, so
$(c,v_2)$ remains and has descendant set $X_2$.
The two-color search cuts only the second input, so it cannot
alter either conclusion. The fresh marker on the retained side
is outside $X_2$.

The set $X_2$ and the label set on the original root side of $c$
are both nonempty. Hence the indicated edge survives in $R$.
In a reduced rooted binary tree, two distinct edges cannot have
the same nonempty descendant set: their subtrees would be nested
with no labeled branch between them, leaving a vertex that
should have been suppressed. This proves uniqueness. A traversal
computing descendant label sets finds the edge in polynomial time.
\end{proof}

\subsection{Deferred binary separation}
\label{subsec:deferred-binary}

We combine the easy and difficult constructions into a
single procedure.
Each call separates the inputs but leaves the side instances
unsolved, so that a second separation can be performed before
the side solvers are called.

The input is a state $\mathcal S=(T,F;b,\mathcal R)$ and
a connector $\bar e$ in the current tree $T$.
Let $W$ be the full current label set, let $Y$ be the
descendant label set of $\bar e$, and put $Z=W\setminus Y$.
These sets include all current labels, including inherited
markers.
For the second separation,
Lemma~\ref{lem:current-connector-representative} identifies
the required connector.

The procedure uses the following constructions.
\begin{enumerate}[(i)]
  \item If $F$ admits $(Y,Z)$, delete $\bar e$ from $T$
        and use direct separation.
        No marker or record is added, and the budget
        remains $b$.

  \item Otherwise, if there is an edge $f$ such that every
        component of $F\setminus\{f\}$ has all its labels in
        $Y$ or all in $Z$, choose any one such edge.
        Use the parallel orientation if all labels below $f$
        lie in $Y$, and the opposite orientation if they lie
        in $Z$. Apply the one-edge easy construction with
        the chosen edge and orientation.
        Append its reconstruction record to $\mathcal R$
        and change the budget to $b+1$.

  \item If neither easy case applies, use the procedure
        of Section~\ref{subsec:difficult-interface}.
        It runs one two-color search without markers and
        one in the parallel orientation for each edge of $F$.
\end{enumerate}
Whenever fresh markers $\delta,\gamma$ are introduced,
the prescribed sides become
$Y^+=Y\cup\{\delta\}$ and $Z^+=Z\cup\{\gamma\}$.
Each terminal state is returned with its remaining budget
and reconstruction records.
We call this procedure \emph{deferred binary separation}.

\begin{lemma}
\label{lem:deferred-binary-interface}
Let $\mathcal S=(T,F;b,\mathcal R)$ be a tree--forest
state on $W$, and let $\bar e$ induce the bipartition
$(Y,Z)$.
The deferred procedure returns candidate terminal pairs
$(T',F')$, each with updated records and a nonnegative
remaining budget $b'$, satisfying the following properties.
\begin{enumerate}[(i)]
  \item In every candidate, $T'$ has two components
        with the prescribed label sets, and every
        component of $F'$ lies entirely on one side.
        The prescribed sets are $Y,Z$ without fresh
        markers and $Y^+,Z^+$ when a record is created.

 \item If $G$ is an \af for $(T,F)$ with
        $\ord(G)\leq\ord(F)+b$, then at least one
        candidate preserves a target $G'$ satisfying
        $\ord(G')\leq\ord(F')+b'$.
        If no record is created, then $G'=G$.
        Otherwise, the target is transformed as follows:
        \begin{itemize}
          \item if $\bar e\in E_T(G)$, its order
                increases by one and both fresh
                marker edges occur in $G'$;
          \item if $\bar e\notin E_T(G)$, its order
                increases by two and both fresh
                markers are singleton components.
        \end{itemize}

  \item Each candidate creates at most one new record.
        If $i$ is the number of budget-charged deletions
        made by the call, then
        \[
          b'=
          \begin{cases}
            b-i,   & \text{if no record is created},\\
            b+1-i, & \text{if one record is created}.
          \end{cases}
        \]
        All records already on $\mathcal R$ are retained.

  \item Fix one initial construction for this binary
        separation, including the edge and orientation
        when markers are introduced.
        Let $\mathcal C_i$ be the collection of returned
        terminal calls reached after exactly $i$ charged
        deletions, counted with multiplicity.
        Then $\sum_i|\mathcal C_i|\varphi^{-i}\leq1$.
        There are $O(|W|)$ available initial constructions.
\end{enumerate}
\end{lemma}

\begin{proof}
Parts~(i), (iii), and~(iv) concern the procedure itself
and require no feasible target.
For part~(ii), assume that $G$ satisfies the stated
feasibility inequality.
For direct separation, parts~(i) and~(ii) follow from
Lemma~\ref{lem:direct-easy-separation}.
For the one-edge easy case, they follow from
Lemmas~\ref{lem:split-marked-target}
and~\ref{lem:singleton-marker-target}.
For the difficult case, they follow from
Lemma~\ref{lem:marked-connector-guess} and the separation
and target-preservation guarantees of
Lemmas~\ref{lem:property of 2LS-Process-1}
and~\ref{lem:2LS-Process-1}.
The target feasibility bound and part~(iii) follow from
the budget rules of
Section~\ref{subsec:separation-budgets} and
Lemma~\ref{lem:temporary-budget-invariant}.

For each fixed initial construction in the difficult case,
part~(iv) follows from
Lemma~\ref{lem:2LS-Process-1 branching complexity}.
In either easy case, there is one deterministic output
with $i=0$, so the weighted sum is one.
The difficult case has $1+|E(F)|=O(|W|)$ initial
constructions, one without markers and one with markers
for each edge of $F$.
\end{proof}

The returned budget can be used directly for the side
that undergoes the second separation.
Indeed, let $A,B$ be the full side label sets of a
target-preserving terminal pair $(T',F')$.
The target gaps on the two sides are nonnegative and satisfy
\[\sum_{C\in\{A,B\}}
    \bigl(\ord(G'[C])-\ord(F'[C])\bigr)
  =\ord(G')-\ord(F')
  \leq b'.\]
Hence the target on either side is feasible with budget $b'$.
The algorithm sets one side aside and applies the second
separation to the other, carrying forward the returned
budget and all reconstruction records.
No allocation of the budget between the two sides
needs to be guessed.

\subsection{Three-way balanced separation}

The two stages produce three sides, but these sides contain fresh
markers as well as labels from the input. The earlier balance bound
applies before these fresh markers are added. We must therefore
bound the extra target components that they can create.
We use \emph{inherited labels} for all labels present at the start
of the current call. These include original taxa and any markers
from outer calls. The two separations introduce \emph{fresh markers}.
Only the fresh markers are counted in the extra term below.

\begin{corollary}
\label{cor:marked-side-gap-bound}
Let $(T,F)$ be the tree--forest input on $X$ at the start
of the current call, before either binary separation,
and let $G$ be an \af for this pair.
Let $E\subseteq E(T)$ be the critical ee-set supplied by
Lemma~\ref{lem:target-witness}, with its guaranteed ordering.
Let $r=|E|\geq3$, and let $c$ be a balanced center for $E$.
For $i\in\{0,1,2\}$, let $X_i$ be the label set of the
corresponding component of $T-c$, with $X_0$ on the root
side, and let $M_i=M_i(c,E)$.
The sets $X_i$ contain the labels present before the two
separations, including any inherited markers.
Follow a target-compatible sequence of separation steps
through both stages.
After both stages, isolate in each side input every fresh
marker that is a singleton in the transformed target.
Remove these singleton markers from the inputs and the target.
Let $(\widetilde T_i,\widetilde F_i)$ and
$\widetilde G_i$ denote the resulting side pair and target.
Then
\[
  \ord(\widetilde G_i)
    -\sigma_{\max}(\widetilde T_i,\widetilde F_i)
  \leq
  \begin{cases}
    M_0+1, & i=0,\\
    M_i,   & i\in\{1,2\}.
  \end{cases}
\]
In particular, every resulting side target has gap at most
$\lfloor r/2\rfloor+1$.
\end{corollary}

\begin{proof}
We first determine which fresh markers can occur on each
side.
If the first stage creates a record, denote its markers by
$\delta_1,\gamma_1$.
The marker $\delta_1$ lies on the separated $X_1$ side.
By Lemma~\ref{lem:current-connector-representative},
the second connector has descendant label set exactly
$X_2$ in the remaining tree.
Thus $\gamma_1$ lies on the final root side.
If the second stage creates a record, its marker
$\delta_2$ lies on the $X_2$ side and $\gamma_2$ lies
on the root side.
Consequently, each child side contains at most one fresh
marker, and the root side contains at most two.

The two connector cuts separate $X_0,X_1,X_2$ without
splitting a target component within any one of these sets.
All other unmarked deletions preserve the current target.
Adding markers and removing singleton fresh markers also
leave the target restrictions to these sets unchanged.
Hence $\widetilde G_i[X_i]=G[X_i]$ for $i\in\{0,1,2\}$.

Let $h_i$ be the number of components of $\widetilde G_i$
containing no label from $X_i$.
Such a component contains only fresh markers.
Since all singleton fresh markers have been removed,
it contains at least two markers.
The marker counts above therefore give
$h_1=h_2=0$ and $h_0 \leq 1$.
The components containing labels from $X_i$ correspond
exactly to the components of $G[X_i]$, so
$\ord(\widetilde G_i)=\ord(G[X_i])+h_i$.

The restriction $\widetilde F_i[X_i]$ is an
$X_i$-subforest of $F[X_i]$.
Therefore
$\ord(\widetilde F_i)
  \geq\ord(\widetilde F_i[X_i])
  \geq\ord(F[X_i])$.
Also, $\widetilde T_i$ is a nonempty tree:
the set $X_i$ is nonempty, and preparation removes only
fresh marker leaves.
It follows that
$\sigma_{\max}(\widetilde T_i,\widetilde F_i)
=\ord(\widetilde F_i)$.

By Corollary~\ref{cor:balanced-side-parameter},
$\ord(G[X_i])-\ord(F[X_i])\leq M_i$.
Combining these bounds gives
\[
\begin{aligned}
  \ord(\widetilde G_i)
    -\sigma_{\max}(\widetilde T_i,\widetilde F_i)
  =\ord(\widetilde G_i)-\ord(\widetilde F_i)
  \leq\ord(G[X_i])-\ord(F[X_i])+h_i \leq M_i+h_i.
\end{aligned}
\]
The bounds on $h_i$ prove the stated inequalities.
Finally, $M_i\leq\lfloor r/2\rfloor$ because $c$ is
balanced, which gives the uniform bound.
\end{proof}

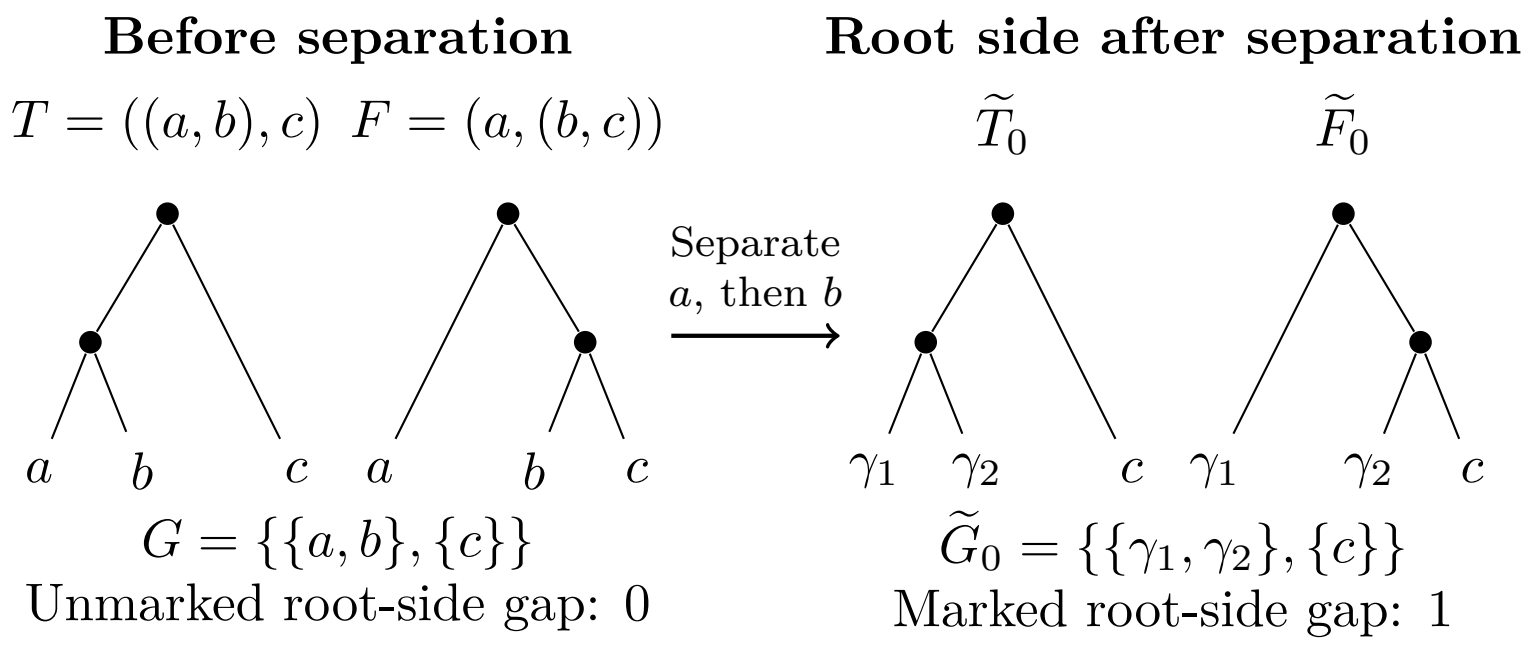
\begin{figure}[t]
\centering
\begin{tikzpicture}[
  scale=0.85,
  every node/.style={font=\small},
  dot/.style={circle,fill,inner sep=1.5pt}
]
\node[font=\small\bfseries] at (2.325,3.55)
  {Before separation};
\node[font=\small\bfseries] at (8.825,3.55)
  {Root side after separation};

\node at (1,2.9) {$T=((a,b),c)$};
\node[dot] (rT) at (1,2.2) {};
\node[dot] (uT) at (0.4,1.2) {};
\node (aT) at (0,0.2) {$a$};
\node (bT) at (0.8,0.2) {$b$};
\node (cT) at (2,0.2) {$c$};
\draw (rT)--(uT) (rT)--(cT)
      (uT)--(aT) (uT)--(bT);

\begin{scope}[xshift=2.65cm]
  \node at (1,2.9) {$F=(a,(b,c))$};
  \node[dot] (rF) at (1,2.2) {};
  \node[dot] (uF) at (1.6,1.2) {};
  \node (aF) at (0,0.2) {$a$};
  \node (bF) at (1.2,0.2) {$b$};
  \node (cF) at (2,0.2) {$c$};
  \draw (rF)--(aF) (rF)--(uF)
        (uF)--(bF) (uF)--(cF);
\end{scope}

\draw[->,thick] (4.92,1.25)--(6.23,1.25)
  node[midway,above,align=center,font=\scriptsize]
  {Separate\\$a$, then $b$};

\begin{scope}[xshift=6.5cm]
  \node at (1,2.9) {$\widetilde T_0$};
  \node[dot] (rTzero) at (1,2.2) {};
  \node[dot] (uTzero) at (0.4,1.2) {};
  \node (gTone) at (0,0.2) {$\gamma_1$};
  \node (gTtwo) at (0.8,0.2) {$\gamma_2$};
  \node (cTzero) at (2,0.2) {$c$};
  \draw (rTzero)--(uTzero) (rTzero)--(cTzero)
        (uTzero)--(gTone) (uTzero)--(gTtwo);
\end{scope}

\begin{scope}[xshift=9.15cm]
  \node at (1,2.9) {$\widetilde F_0$};
  \node[dot] (rFzero) at (1,2.2) {};
  \node[dot] (uFzero) at (1.6,1.2) {};
  \node (gFone) at (0,0.2) {$\gamma_1$};
  \node (gFtwo) at (1.2,0.2) {$\gamma_2$};
  \node (cFzero) at (2,0.2) {$c$};
  \draw (rFzero)--(gFone) (rFzero)--(uFzero)
        (uFzero)--(gFtwo) (uFzero)--(cFzero);
\end{scope}

\node[align=center] at (2.325,-0.6)
  {$G=\{\{a,b\},\{c\}\}$\\
   Unmarked root-side gap: $0$};
\node[align=center] at (8.825,-0.6)
  {$\widetilde G_0=\{\{\gamma_1,\gamma_2\},\{c\}\}$\\
   Marked root-side gap: $1$};
\end{tikzpicture}

\caption{The original trees and target are on the left. The root-side
trees and target after separating $a$ and then $b$ are on the right.
The component $\{\gamma_1,\gamma_2\}$ survives singleton stripping,
so the root-side gap increases from zero to one.}
\label{fig:marker-only-component}
\end{figure}

Figure~\ref{fig:marker-only-component} shows how a component
containing only fresh markers can increase the root-side gap.
Take $T=((a,b),c)$ and $F=(a,(b,c))$, and let $G$ have components
with label sets $\{a,b\}$ and $\{c\}$. At the parent of $a$ and
$b$ in $T$, the unmarked root side has label set $X_0=\{c\}$.
Both $G[X_0]$ and $F[X_0]$ have one component, so the gap on
this side is zero before marking.

Now separate $a$ and then $b$ using parallel marker operations.
The first operation leaves a target component with labels
$\{b,\gamma_1\}$ on the remaining side. The second replaces
$b$ in this component by $\gamma_2$, leaving
$\{\gamma_1,\gamma_2\}$ on the root side. The two markers
survive singleton stripping because they belong to the same
component. This component gives $h_0=1$ in the proof above.
Together with $\{c\}$, the root-side target has two components,
while the second input $\widetilde F_0$ is a single tree.
The resulting gap is $2-1=1$.
The full algorithm handles this three-label instance in its
initial constant-budget call.

Let $\mathcal A$ be a tree--forest solver that, given
$(T,F;\kappa)$, returns an explicit \maf if
$m(T,F)\leq\ord(F)+\kappa$, and returns \FAIL{} otherwise.
The solver treats all input labels, including markers,
as ordinary leaf labels.
We call such a solver \emph{compatible}.
The auxiliary solver \algname{Alg-MAF$_M$} of
Section~\ref{sec:auxiliary-ls-solver} is compatible.

For the running-time analysis, we number the initial
constructions of deferred binary separation described in
Section~\ref{subsec:deferred-binary}.
At a given separation state and connector, each available
initial construction is called a \emph{choice}.
In either easy case, the procedure selects a single
construction, which is assigned index one.

For a difficult state, fix an ordering
$f_1,\ldots,f_{|E(F)|}$ of the edges of the current
second forest $F$.
The choices have the following indices.
\[
  \begin{array}{c|l}
    \text{index} & \text{initial construction}\\ \hline
    1
      & \text{unmarked construction}\\
    a+1
      & \text{marked construction at }f_a
        \text{ with parallel orientation}
  \end{array}
  \qquad (1\leq a\leq |E(F)|).
\]
For a fixed choice, the subsequent two-color process
may branch repeatedly and return several terminal states.
All these states arise from the same initial construction
and are therefore counted under that choice.

Let $n$ be the number of labels at the start of the current
balanced-center call, including inherited markers.
At most four fresh markers are introduced at this level.
Every current forest therefore has at most $n+4$ labels
and $O(n)$ edges, so each choice list has length $O(n)$.

We group the two-stage searches by pairs of choice indices
$(j_1,j_2)$.
The first separation uses choice $j_1$.
For each terminal state returned by that separation,
the second separation uses choice $j_2$ from the list
available in that state.
The second-stage lists may differ because the first
separation can return different second forests.

Thus $j_2$ refers to a position in the list for the current
second-stage state. For $j_2=1$, an easy state uses its only
construction, either direct separation or one-edge marked
separation. A difficult state uses its unmarked construction.
For $j_2=2$, a difficult state uses the parallel marked
construction at the first edge in its own ordering of $E(F)$.
An easy state has no second entry and produces no candidates
for this index. In general, an index beyond the end of the
current list produces no candidates.

Let $J=O(n)$ be a common upper bound on the lengths of
the choice lists. For each center, we consider index pairs
$(j_1,j_2)$ with $1\leq j_1,j_2\leq J$, giving $O(n^2)$ pairs.
For each fixed pair, we include all terminal calls
generated by the corresponding two-stage searches.
Calls reached through different search paths are counted
separately, even when their forests and remaining budgets
coincide.

\begin{lemma}
\label{lem:balanced-center-interface}
Let $(T,F;k)$ be a tree--forest instance with $k\geq0$.
The two-stage separation construction has the following
properties.
\begin{enumerate}[(i)]
  \item Suppose that $G$ is an \af for $(T,F)$ with
        $3\leq r
          =\ord(G)-\sigma_{\max}(T,F)
          \leq k$.
        Among the enumerated centers and separation
        choices, there is a target-compatible construction
        producing three independent sides with inherited
        label sets $X_0,X_1,X_2$ and at most two new
        reconstruction records.

        Let $b$ be its final temporary budget.
        Choose for singleton preparation exactly the fresh
        markers that are singletons in the final transformed
        target.
        Every prepared side target then has gap at most
        $\min\{\lfloor r/2\rfloor+1,b\}$.
        Compatible side solvers with budget
        $\kappa=\min\{\lfloor k/2\rfloor+1,b\}$
        return forests that reconstruct to an \af $H$
        for $(T,F)$ with $\ord(H)\leq\ord(G)$.

  \item Fix any nonroot internal center and any pair
        of local construction indices $(j_1,j_2)$.
        Let $\mathcal D_i$ be the collection of terminal
        separation calls reached after exactly $i$ charged
        deletions across both stages, before singleton
        preparation.
        Calls are counted with multiplicity.
        Then $\sum_i|\mathcal D_i| \cdot \varphi^{-i} \leq 1$.
        A terminal candidate with $t$ new records has
        remaining budget $b=k+t-i$ with $0\leq t\leq 2$.
        These bounds hold whether or not the original
        instance is feasible for budget $k$.
        The enumeration of centers and construction
        indices adds only a polynomial factor.
\end{enumerate}
\end{lemma}

\begin{proof}
For part~(i), fix an \af $G$ with
$3\leq r=\ord(G)-\sigma_{\max}(T,F)\leq k$.
Lemma~\ref{lem:target-witness} gives an ordered critical
ee-set $E\subseteq E(T)$ of size $r$ for $G$.
By Lemma~\ref{lem:balanced-center-generic}, there is a
balanced nonroot center $c$ for $E$.
This vertex is among the enumerated centers.
Let $X_0,X_1,X_2$ be its three side label sets, with
$X_0$ on the root side.

Let $e_1,e_2$ be the two child connectors of $c$.
Apply Lemma~\ref{lem:deferred-binary-interface}(ii)
at $e_1$ and choose a target-preserving terminal state.
Set its child side aside, and let $b_1$ be the returned
budget.
The nonnegative-gap identity following that lemma shows
that the target on the remaining side is feasible for
$b_1$.
By Lemma~\ref{lem:current-connector-representative},
the second connector has a current representative whose
descendant label set is $X_2$.
A second application of
Lemma~\ref{lem:deferred-binary-interface}(ii), with budget
$b_1$, therefore gives a target-preserving result.
The two stages produce three independent sides with
inherited label sets $X_0,X_1,X_2$.
Each stage creates at most one record.

Let $G'$ be the final transformed target, let $F'$ be
the union of the three second-input forests, and let $b$
be the final temporary budget.
Choose for singleton preparation exactly the fresh markers
that are singleton components of $G'$.
This subset is among those enumerated by the algorithm.
Corollary~\ref{cor:marked-side-gap-bound} bounds every
prepared side target gap by $\lfloor r/2\rfloor+1$.

There is also a bound by $b$.
Lemma~\ref{lem:temporary-budget-invariant} gives
$\ord(G')-\ord(F')\leq b$.
Since the sides are independent and each first input
is a tree, the left-hand side is the sum of three
nonnegative target gaps.
Each side target gap is therefore at most $b$.
Every side contains the nonempty inherited set $X_i$,
so Lemma~\ref{lem:singleton-preparation-reconstruction}
applies.
Its part~(i) shows that preparation cannot increase
these gaps.
Thus every prepared side target has gap at most
$\min\{\lfloor r/2\rfloor+1,b\}$.
Since $r\leq k$, the supplied budget
$\kappa=\min\{\lfloor k/2\rfloor+1,b\}$ is sufficient
for every side.
Compatible solvers return \mafs for the prepared pairs.
Part~(ii) of
Lemma~\ref{lem:singleton-preparation-reconstruction}
then gives a reconstructed \af $H$ with
$\ord(H)\leq\ord(G)$.
This proves part~(i).

For part~(ii), fix any nonroot internal center and
any pair of local construction indices $(j_1,j_2)$.
Let $\mathcal C_{i_1}$ be the first-stage terminal calls
reached after $i_1$ charged deletions under choice $j_1$.
For each such call $s$, let $\mathcal C_{i_2}(s)$ be
the second-stage terminal calls reached after $i_2$
further charged deletions under its local choice $j_2$.
All calls are counted with multiplicity.
If a specified choice is absent, the corresponding
collection is empty.

Lemma~\ref{lem:deferred-binary-interface}(iv) applies
to the first-stage search and to each second-stage search,
without assuming that a feasible target exists.
Consequently,
\[
\begin{aligned}
  \sum_i|\mathcal D_i| \cdot \varphi^{-i} =
    \sum_{i_1}\sum_{s\in\mathcal C_{i_1}}
      \varphi^{-i_1}
      \sum_{i_2}|\mathcal C_{i_2}(s)| \cdot \varphi^{-i_2} \leq
    \sum_{i_1}|\mathcal C_{i_1}| \cdot \varphi^{-i_1} \leq1.
\end{aligned}
\]

The temporary budget starts at $k$.
Each new record increases it by one, and each charged
deletion decreases it by one.
Hence every terminal candidate reached after $i$
charged deletions and $t$ new records satisfies
$b=k+t-i$.
Each stage creates at most one record, so $0\leq t\leq2$.
These budget and counting bounds hold for every branch,
regardless of whether the input is feasible for $k$.

Finally, there are $O(n)$ centers and $O(n)$ local
construction indices in each stage.
Their enumeration therefore contributes only an
$O(n^3)$ factor.
\end{proof}

\subsection{The algorithm and its running time}

Algorithm~\ref{alg:BC-Alg-MAF_3} gives the balanced-center algorithm
\algname{BC-Alg-MAF$_3$} for tree--forest instances $(T,F;k)$.
It tries every candidate center and
both separation stages before making any side call. Each state
carries its reconstruction stack $\mathcal R$, even when it is
not shown. The descendant set for the second connector is stored
before the first separation.
The budget is shared across the two stages:
$b=k+t-i$,
where $t$ is the number of records created at this level and $i$
is the total number of charged deletions at this level. Neither
count is reset after the first separation. The total forest order
includes the side already set aside.

The algorithm does not know the target $G$ or its singleton
markers. It therefore tries every subset of the fresh markers.
There are at most four such markers, so there are at most $16$
subsets. A failed side call rejects only the current singleton
subset.

\begin{algorithm}[htbp]
\caption{\algname{BC-Alg-MAF$_3$}$(T,F;k)$ with side solver $\mathcal A$}
\label{alg:BC-Alg-MAF_3}
\small

\textbf{Input:}
A nonempty tree--forest pair $(T,F)$, an integer budget $k$,
and a compatible side solver $\mathcal A$.\\
\textbf{Output:}
An explicit \maf for $(T,F)$ if $m(T,F)\leq\ord(F)+k$,
and \FAIL{} otherwise.

\medskip
\begin{algorithmic}[1]
\If{$k<0$}
  \State \Return \FAIL{}.
\EndIf
\State $k\gets\min\{k,|L(T)|-\ord(F)\}$;
\State $H\gets\algname{Alg-MAF$_M$}(T,F;\min\{k,2\})$;
\If{$H\ne\FAIL$}
  \State \Return $H$.
\EndIf
\If{$k\leq2$}
  \State \Return \FAIL{}.
\EndIf

\State $G_{\mathrm{best}}\gets\bot$;
\For{each nonroot internal vertex $c$ of $T$}
  \State store its child connectors $e_1,e_2$ and their
    descendant label sets $X_1,X_2$;
  \State apply deferred binary separation to $(T,F;k,\varnothing)$
    at $e_1$, obtaining a list $\mathcal C_1$;
  \For{each state in $\mathcal C_1$}
    \State set aside the side containing $X_1$;
    \State form $\mathcal S$ from the remaining side, the returned
      budget, and the returned records;
    \State find the current representative $\bar e_2$ of $e_2$
      by Lemma~\ref{lem:current-connector-representative};
    \State apply deferred binary separation to $\mathcal S$ at
      $\bar e_2$, obtaining a list $\mathcal C_2$;
    \For{each state in $\mathcal C_2$}
      \State use the two returned sides and the side set aside
        as the three side pairs;
      \State let $b$ be the final budget and $M$ the set of fresh
        markers from both stages;
      \For{each subset $S\subseteq M$}
        \State copy the three side pairs and isolate the markers
          in $S$ in both inputs;
        \State strip the isolated markers to obtain
          $(\widetilde T_i,\widetilde F_i)$ for $i=0,1,2$;
        \State $\kappa\gets\min\{\lfloor k/2\rfloor+1,b\}$;
        \State $H_i\gets\mathcal A(\widetilde T_i,\widetilde F_i;\kappa)$
          for $i=0,1,2$;
        \If{any $H_i$ equals \FAIL{}}
          \State continue with the next subset $S$;
        \EndIf
        \State restore the stripped markers in their side
          solutions as singletons;
        \State let $H$ be the union of the three restored side solutions;
        \State apply this level's reconstruction records to $H$
          in reverse order;
        \If{$H$ is an \af for $(T,F)$ and $\ord(H)\leq\ord(F)+k$}
          \State Set $G_{\mathrm{best}}\gets H$ if
            $G_{\mathrm{best}}=\bot$ or $\ord(H)<\ord(G_{\mathrm{best}})$.
        \EndIf
      \EndFor
    \EndFor
  \EndFor
\EndFor
\State \Return \FAIL{} if $G_{\mathrm{best}}=\bot$,
  and $G_{\mathrm{best}}$ otherwise.
\end{algorithmic}
\end{algorithm}

The running-time improvement depends on the exponential base of
the side solver, rather than on its particular branching rules.
We state the resulting solver transformation first and then
substitute the auxiliary solver's bound.

\begin{theorem}
\label{thm:BC-Maf}
Let $\mathcal A$ be a compatible tree--forest solver with running
time $O^*(B^\kappa)$ for a fixed constant $B>\varphi=(1+\sqrt5)/2$.
With $\mathcal A$ as its side solver,
$\algname{BC-Alg-MAF$_3$}$ returns an explicit \maf for $(T,F)$
if $\op_3(T,F)\leq k$, and returns \FAIL{} otherwise.
Its running time is $O^*((\sqrt{\varphi B})^k)$, and the
resulting solver is compatible.
\end{theorem}

\begin{proof}
Let $G$ be an \maf and put $r=\op_3(T,F)$. If $r\leq k$ and
$r\leq2$, the initial call returns an \maf, since
$\op_2(T,F)=\op_3(T,F)$. If $3\leq r\leq k$, that
constant-budget call fails and the algorithm enters the separation
phase. Lemma~\ref{lem:balanced-center-interface} supplies a center and
a target-compatible candidate whose reconstructed forest has order at most
$\ord(G)$. That forest is therefore an \maf. The algorithm
retains a candidate of minimum order. Every returned candidate is
checked against the original inputs and the threshold
$\ord(F)+k$, so no forest is returned when $r>k$.

The resulting algorithm is a compatible tree--forest solver.
Every returned forest contains all labels present at the
start of the call, including inherited markers.
Reconstruction removes only the fresh markers introduced
by this call.

For the running time, fix a center and a pair of local
choice indices $(j_1,j_2)$. Let $\mathcal D_i$ contain the terminal calls with
$i$ charged deletions in total, counted with multiplicity. At most
two records are created, so their temporary budgets satisfy $b\leq k+2-i$.
Every side-call budget therefore satisfies
$\kappa\leq\min\{k/2+1,k+2-i\}$.
The weighted bound in
Lemma~\ref{lem:balanced-center-interface} gives
\begin{align*}
  \sum_i|\mathcal D_i| \cdot B^{\min\{k/2+1,k+2-i\}}
  &\leq \left(\sum_i|\mathcal D_i| \cdot \varphi^{-i}\right)
    \max_i\left(\varphi^i \cdot
      B^{\min\{k/2+1,k+2-i\}}\right)\\
  &\leq \varphi B \cdot (\varphi B)^{k/2}.
\end{align*}
For $i\leq k/2+1$, use the bound $k/2+1$ and
$\varphi^i\leq\varphi^{k/2+1}$. For $i\geq k/2+1$, use
$k+2-i$ and
\[
  \varphi^i B^{k+2-i}
  =\varphi B \cdot (\varphi B)^{k/2} \cdot (\varphi/B)^{i-k/2-1}
  \leq \varphi B \cdot (\varphi B)^{k/2}.
\]

Each terminal state generates at most three side calls for
each of at most $16$ subsets. Thus, for the fixed center and
index pair, the total time spent in the side solvers is
$O^*((\varphi B)^{k/2})$.

We now bound the work of generating the separation states for
the same center and index pair. The first separation starts
its two-color search with budget at most $k+1$, including any
increase from its marker operation. By
Lemma~\ref{lem:2LS-Process-1 branching complexity}, this stage
takes $O^*(\varphi^{k+1})$ time. The easy constructions take
polynomial time.

Let $\mathcal C_{i_1}$ be the collection of first-stage terminal
states reached under choice $j_1$ after $i_1$ charged deletions,
counted with multiplicity. The budget returned with each such
state is at most $k+1-i_1$. Allowing for a marker operation in
the second separation, the search for choice $j_2$ starts with
budget at most $k+2-i_1$. Its cost is therefore bounded by
$O^*(\varphi^{k+2-i_1})$ per first-stage state, including the
polynomial work when the second separation is easy or the
index is absent. Summing over the first-stage states gives
$O^*(\varphi^{k+2} \cdot \sum_{i_1}|\mathcal C_{i_1}| \cdot \varphi^{-i_1})$.
The sum is at most one by
Lemma~\ref{lem:deferred-binary-interface}(iv), so both
separations together take $O^*(\varphi^k)$ time.

Preparing the side inputs, reconstructing the combined forest,
and checking it against the input pair take polynomial time per
terminal state and subset. To verify an \af, compute the minimal
subtree for each component label set, check disjointness in
each input, and compare the reduced labeled trees. Since
$i\leq k+2$, the weighted bound also gives
$\sum_i|\mathcal D_i|\leq\varphi^{k+2}$. All work outside the
side calls therefore takes $O^*(\varphi^k)$ time for the fixed
center and index pair.

Since $\varphi^k\leq(\varphi B)^{k/2}$, the side calls and
the remaining work together take $O^*((\varphi B)^{k/2})$
time for each center and index pair. Summing over $O(n)$
centers and $O(n^2)$ index pairs contributes a polynomial
factor. The initial constant-budget call also takes polynomial
time. The total running time is therefore
$O^*((\sqrt{\varphi B})^k)$.
\end{proof}

\begin{corollary}
\label{cor:first-composition}
With side solver $\algname{Alg-MAF$_M$}$,
$\algname{BC-Alg-MAF$_3$}$ is a compatible optimization solver
with running time $O^*(1.9455^k)$.
\end{corollary}

\begin{proof}
The auxiliary solver has base $B=2.3391>\varphi$. Apply
Theorem~\ref{thm:BC-Maf} and use
$\sqrt{\varphi\cdot2.3391}=1.945441\ldots<1.9455$.
\end{proof}

\subsection{Recursive composition}

The algorithm above takes the same type of input as its side solver and gives the same output guarantee. We can therefore replace the side solver with this algorithm and apply the framework again.

Let $\mathcal A_0$ denote \algname{Alg-MAF$_M$}, and put
$\beta_0=2.3391$.
For $j\geq0$, define $\mathcal A_{j+1}$ by
Algorithm~\ref{alg:BC-Alg-MAF_3}, with $\mathcal A_j$ as its
side solver. The initial constant-budget call is always
\algname{Alg-MAF$_M$}.
The composition depth fixes which solver is used: a call to
$\mathcal A_{j+1}$ calls only $\mathcal A_j$ on its sides,
and $\mathcal A_0$ invokes the auxiliary solver directly.
Moreover, separation is reached only when $k\geq3$, and its
side budgets are at most $\lfloor k/2\rfloor+1<k$.
Thus every side call also has a smaller supplied budget.
The initial call handles all instances with at most two labels,
whose optimum gap is at most one. If $k\leq2$ and that call
fails, the algorithm returns \FAIL{} immediately.

After the selected common singletons are stripped, every side
input is a tree--forest pair, so its refined parameter and
additional-cut parameter agree. Markers inherited from an outer
call are treated as ordinary labels by the inner solver. Only
the markers selected to be singleton are removed before that
call and restored as singletons afterward.

An inner call removes all markers created by its own records
before it returns. It retains every inherited marker in its input.
Thus, after the stripped markers have been restored, the returned
forest contains every marker needed by an outer record. The record
can read whether each marker is singleton and, otherwise, its
current neighbor. Reconstruction then applies
Lemma~\ref{lem:maf combination for case 2.2} in reverse record
order.

\begin{theorem}
\label{thm:recursive-composition}
For every fixed integer $x\geq0$, the algorithm $\mathcal A_x$
returns an \maf on a tree--forest input $(T,F;k)$ if
$\op_3(T,F)\leq k$, and returns \FAIL{} otherwise. Let 
$\beta_x=\varphi^{\,1-2^{-x}} \cdot 2.3391^{2^{-x}}$.
Its running time is $O^*(\beta_x^k)$. For each fixed $x$, the
suppressed polynomial factor has fixed degree; its degree and
hidden constant may depend on $x$.
In particular, if
$x=x(\varepsilon)$ is fixed to obtain base at most
$\varphi+\varepsilon$, both the polynomial degree and the hidden
constant may depend on $\varepsilon$.
\end{theorem}

\begin{proof}
The auxiliary solver $\mathcal A_0$ is compatible and has base
$\beta_0=2.3391>\varphi$. If $\mathcal A_j$ is compatible with
base $\beta_j>\varphi$, Theorem~\ref{thm:BC-Maf} shows that
$\mathcal A_{j+1}$ is compatible with base
$\beta_{j+1}=\sqrt{\varphi \beta_j}>\varphi$. Induction gives both the
output guarantee and the recurrence. Solving it yields
$\beta_x=\varphi^{1-2^{-x}}2.3391^{2^{-x}}$.

The auxiliary solver has a fixed polynomial running-time factor.
At each composition level, the enumeration and local processing
analyzed in Theorem~\ref{thm:BC-Maf} introduce only further
polynomial factors. Induction on the composition depth therefore
gives a fixed polynomial factor for every fixed $x$.
Along a chain of depth at most $x$, there are
at most $n+4x$ labels. For fixed $x$ and $n\geq1$, replacing this
size by $n$ changes only a constant depending on $x$.
There are also only $O(x)$ active constant-size reconstruction
records along the chain, which are covered by that constant.
\end{proof}

In particular, $\beta_1=1.945441\ldots$, and $\beta_2=\sqrt{\varphi \beta_1}
     =1.774201\ldots<1.7743$.
Two composition levels therefore give the $O^*(1.7743^k)$ bound
in Theorem~\ref{thm:final-2}.
The formula $\beta_x=\varphi(2.3391/\varphi)^{2^{-x}}$
shows that the bases converge to $\varphi$. For a fixed
$\varepsilon>0$, choose a fixed depth satisfying
$2^{-x} \cdot \log(2.3391/\varphi) \leq\log(1+\varepsilon/\varphi)$.
This gives an $O^*((\varphi+\varepsilon)^k)$ algorithm, with
$x=O(\log(1/\varepsilon))$ as $\varepsilon\downarrow0$.
The depth is chosen before the input is processed. Thus the
polynomial degree and the hidden constant may depend on
$\varepsilon$, but not on the input.

If the optimum budget is unknown, run the corresponding algorithm
with budgets $0,1,\ldots$ and stop at the first success. It returns
an \maf, and the geometric sum preserves the exponential base.

\section{Kernelization without the Extra Root Label}
\label{sec:root-kernel-discussion}

Kelk et al.~\cite{kelk2023cyclic} studied the kernelization for the MAF problem on $X$-trees with an extra root labeled $\rho$.
Our algorithms use $X$-trees without this extra label. Before using
their kernelization, we show that its reduction
rules are also safe for our setting. 
We prove this for the subtree, chain, and
$3$-$2$-chain reductions. We then derive a size bound for the
reduced input and apply it to our running-time bounds.

\subsection{Reduction rules on the ordinary labels}

We apply the following rules to the unaugmented pair. All labels
mentioned in the rules are ordinary taxa. We use the chain order
of Section~5 of~\cite{kelk2023cyclic}. Let
$(x_1,\ldots,x_\ell)$ be a sequence of distinct labels, with
$\ell\geq2$, and let $u_i$ be the parent of $x_i$.
The sequence is a \emph{chain} if $u_1=u_2$ or $(u_2,u_1)$
is an edge, and $(u_i,u_{i-1})$ is an edge for every
$3\leq i\leq\ell$.
A chain is pendant when $u_1=u_2$.
\begin{enumerate}[(i)]
  \item {\bf Subtree Reduction:} Replace a maximal common pendant subtree with at least
        two labels by one representative label.
  \item {\bf Chain Reduction:} Shorten a maximal common chain of length at least four
        to its first three labels.
  \item {\bf 3-2-Chain Reduction:} Suppose $(x_1,x_2,x_3)$ is a pendant $3$-chain in one
        tree and $(x_i,x_3)$ is a cherry in the other, where
        $i\in\{1,2\}$. Delete $x_j$ from both trees, where
        $\{x_i,x_j\}=\{x_1,x_2\}$.
\end{enumerate}
Deleted labels are removed by restriction. If the two current trees are isomorphic, we solve
the instance directly.

The rules use the same local configurations as in the cited
kernelization. The next lemma gives their effect on our
unaugmented objective.

\begin{lemma}
\label{lem:kelk-unaugmented-compatibility}
Let $(S_1,S_2)$ be the result of applying one of the
three reductions above to two unaugmented $X$-trees
$(T_1,T_2)$. Then
\[
  m(S_1,S_2)=
  \begin{cases}
    m(T_1,T_2),&\text{for a subtree or chain reduction},\\
    m(T_1,T_2)-1,&\text{for a $3$-$2$-chain reduction}.
  \end{cases}
\]
Given any \maf for $(S_1,S_2)$, we can recover a \maf for
$(T_1,T_2)$ in polynomial time. For a subtree or chain
reduction, the recovered forest has the same number of
components as the given \maf. For a $3$-$2$-chain reduction,
we add the deleted label as a singleton component, increasing
the number of components by one.
\end{lemma}

\begin{proof}
Appendix~\ref{app:kelk-compatibility} proves all three cases
and gives the operations for recovering an \af for the original pair.
All three arguments concern ordinary leaf-labeled trees and
apply also when the configuration meets the original root.
\end{proof}

The lemma justifies applying all three rules without adding
$\rho$. Suppose an exhaustive sequence contains $s$
$3$-$2$-chain reductions and produces $(S_1,S_2)$. Then
$m(T_1,T_2)=m(S_1,S_2)+s$.
Since the refined parameter of a tree pair is its optimum order
minus one, a supplied budget $k$ becomes $K=k-s$.
Thus the original instance is feasible exactly when
$m(S_1,S_2)-1\leq K$. If $K<0$, it is infeasible.
We store the reduction sequence and undo it in reverse order
after solving the reduced pair.

\subsection{Kernel size under our root convention}

The two conventions may give different optimum orders, but the
difference is at most one.

\begin{lemma}
\label{lem:root-convention-comparison}
Let $m_0=m(T_1,T_2)$ and
$m_\rho=m(T_1^\rho,T_2^\rho)$.
Then $m_0\leq m_\rho\leq m_0+1$.
\end{lemma}

\begin{proof}
Take an \af for the augmented pair.
Removing $\rho$ and its incident edge, if present, and applying
forced contraction gives an \af for $(T_1,T_2)$ whose order is
no larger.
Thus $m_0\leq m_\rho$.
Conversely, take an \af of order $m_0$ for $(T_1,T_2)$.
In each augmented input, delete the edge joining $\rho$ to the
original root.
The same \af, together with the singleton component
$\{\rho\}$, is an \af for the augmented pair of order $m_0+1$.
Hence $m_\rho\leq m_0+1$.
\end{proof}

Let $(S_1,S_2)$ be the preprocessed tree pair, let $X$ be its
ordinary label set, and let $K\geq0$ be its remaining budget.
Choose a fresh auxiliary root label $\eta\notin X$ for the
size argument below. In particular, it is distinct from any leaf
label $\rho$ in an encoded rSPR input. The budget $K$ belongs
to this tree pair; recursive side budgets are still denoted
by $b$ or $\kappa$.

Since both inputs are trees, $\sigma_{\max}(S_1,S_2)=1$, and hence
$\op_3(S_1,S_2)=m(S_1,S_2)-1$.
Thus feasibility for budget $K$ means $m(S_1,S_2)-1\leq K$, or
$m(S_1,S_2)\leq K+1$.
We use this inequality to express the kernel size in terms of $K$.

Assume that none of the three reduction rules applies to
$(S_1,S_2)$. Isomorphic pairs are solved directly, and the bound
$|X|\leq9K+6$ holds trivially when $|X|\leq3$. We therefore
consider a nonisomorphic pair with $|X|\geq4$.

To apply the kernel-size theorem of Kelk et
al.~\cite{kelk2023cyclic}, consider $(S_1^\eta,S_2^\eta)$, obtained
by adding the extra root label. Its full label set is
$X\cup\{\eta\}$. Kelk et al. also use $X$ for the ordinary taxa:
their \afs partition $X\cup\{\eta\}$, whereas their
kernel-size bound counts only the labels in $X$.

We first check that the augmented pair still admits none of the
three reductions. Adding $\eta$ above the root does not change
any parent relation between ordinary vertices. Since chains
contain only ordinary taxa, it creates no new chain or
$3$-$2$-chain configuration. It also creates no new pendant
subtree on a proper subset of $X$. The only additional pendant
subtree is the whole original tree, which could be common to
both inputs only if $S_1$ and $S_2$ were isomorphic. Thus the
augmented pair is irreducible under the cited rules.

Put $d=d_{\rm rSPR}(S_1^\eta,S_2^\eta)$. The pair is
nonisomorphic, so $d\geq1$. Theorem~6.2 of
Kelk et al.~\cite{kelk2023cyclic} gives $|X|\leq9d-3$.
For an unaugmented instance feasible for $K$,
Lemma~\ref{lem:root-convention-comparison} and the feasibility
inequality above give
$d = m(S_1^\eta,S_2^\eta)-1
    \leq m(S_1,S_2)
    \leq K+1$.
Therefore, $|X|\leq9(K+1)-3=9K+6$.
After exhaustive reduction, a nonisomorphic pair with more than
$9K+6$ ordinary taxa is therefore infeasible for budget $K$ and
can be rejected. The algorithm continues on $(S_1,S_2)$;
$\eta$ is added only in the proof of this size bound.

\subsection{Recursive sizes and running time}

The preprocessing is performed once, on the original tree pair.
Recursive calls inherit its ordinary labels. Edge deletion,
restriction, and decomposition do not introduce new ordinary
taxa. Each fixed composition level adds only a constant number
of marker labels.

\begin{lemma}
\label{lem:kernel-inheritance}
After exhaustive preprocessing and the size test, every
tree--forest instance arising within at most $x$ nested
applications of the framework has at most
$9K+6+O(x)$ labels. For fixed $x$, this is $O(K+1)$.
\end{lemma}

\begin{proof}
Every ordinary label belongs to the reduced original label set,
which has size at most $9K+6$. The only added labels are markers,
and at most $O(x)$ occur along a chain of $x$ nested applications.
Adding these counts proves the bound.
\end{proof}

This argument uses no kernelization rule on a recursive
tree--forest state. The initial reduction and the inherited label
bound are sufficient.

\begin{lemma}
\label{lem:cubic-kernelization}
For a tree pair on $n$ ordinary taxa, we can apply the three
reduction rules exhaustively and perform the size test in
$O(n^3)$ time. Given a \maf for the reduced pair, we can
recover a \maf for the original pair in $O(n^3)$ time.
\end{lemma}

\begin{proof}
A current tree on $m$ taxa has $O(m)$ vertices and edges.
Canonical rooted subtree signatures identify common pendant
subtrees. Parent pointers identify chain continuations and
pendant $3$-$2$ configurations. Comparing the candidate chains
and their label orders in the two trees, together with rebuilding
the signatures and pointers, takes $O(m^2)$ time by direct scans.

Every successful reduction removes at least one ordinary taxon,
so there are fewer than $n$ successful reductions. Repeating a
scan after each one, followed by a final scan, costs $O(n^3)$.
We store the type, labels, and attachment information for each
reduction and process the records in reverse order to recover the
original solution. To undo a subtree reduction, expand the
representative label into the original common subtree. To undo a
$3$-$2$-chain reduction, add the deleted label as a singleton
component. Each of these operations takes $O(n)$ time.

To undo a chain reduction, compute the minimal representation of
each component in each tree. One traversal per component takes
$O(n^2)$ in total. Scan these representations at the two ends
of the retained chain, including components with no chain label
whose representations pass through the chain. For two input trees,
the constructive normalization in
Lemma~\ref{lem:common-chain-normalization} uses a constant
number of splits and merges. At most two splits are needed
in the case that requires repeated splitting. Each label set
update and the restoration of the deleted chain labels take
$O(n)$ time. Thus undoing one chain reduction takes $O(n^2)$
time. Processing all the records to recover the original
solution takes $O(n^3)$ time.
\end{proof}

If the optimum budget is unknown, preprocess once and retain
the number $s$ of $3$-$2$-chain reductions. A trial budget $q$ for the
original pair corresponds to $K=q-s$ on the reduced pair.
Trials with $q<s$ fail immediately. For each later trial, apply
the size test with this $K$ and, if it passes, run the solver.
At the first successful trial, undo the recorded reductions
in reverse order to recover a \maf for the original pair.
The geometric sum over trial budgets preserves the exponential
base.

At each fixed composition depth $x$,
Theorem~\ref{thm:recursive-composition} gives running time
$O^*(\beta_x^k)$ with a fixed polynomial factor. After the one-time
kernelization, this becomes $O(\beta_x^k \cdot (k+1)^{O(1)}+n^3)$,
since the reduced budget is at most the original budget and all
recursive inputs have $O(k+1)$ labels for fixed $x$. The polynomial
degree and hidden constant may depend on $x$.
The kernelization changes the polynomial dependence on the input
size. The exponential bases remain those proved by the
balanced-center analysis.

\begin{corollary}
\label{cor:kernelized-explicit-bounds}
Let $T_1,T_2$ be unaugmented $X$-trees with $|X|=n$, and put $k^*=m(T_1,T_2)-1$. An \maf can be computed in
$O(1.7743^{k^*} \cdot (k^*+1)^{O(1)}+n^3)$
time. More generally, for every fixed $\varepsilon>0$, choose a
fixed depth $x=x(\varepsilon)$ with
$\beta_x\leq\varphi+\varepsilon$. The running time is then
$O((\varphi+\varepsilon)^{k^*} \cdot (k^*+1)^{O(1)}+n^3)$.
The polynomial degree and hidden constant may depend on
$\varepsilon$. The $n^3$ term covers the one-time preprocessing
and the recovery of a \maf for the original pair.
\end{corollary}

\begin{proof}
The reductions preserve the unaugmented optimum with the recorded
offset $s$. Lemma~\ref{lem:kernel-inheritance} bounds the size of
every recursive input, and
Lemma~\ref{lem:cubic-kernelization} bounds the time for
preprocessing and recovering a \maf for the original pair.
Substitute these bounds into
Theorem~\ref{thm:recursive-composition}, using $x=2$ for the first
claim, and use the successive-budget procedure when
$k^*$ is unknown.
\end{proof}

\section{Conclusion}

We developed a balanced divide-and-conquer framework for the
maximum agreement forest problem on two rooted binary phylogenetic
trees. For $k^*=m(T_1,T_2)-1$, two composition levels give an
$O^*(1.7743^{k^*})$-time algorithm. More generally, for every fixed
$\varepsilon>0$, a suitable fixed number of compositions gives
$O^*((\varphi+\varepsilon)^{k^*})$ time, where
$\varphi=(1+\sqrt5)/2$. The polynomial degree and hidden constant
may depend on $\varepsilon$. The compatibility of the reduction
rules with our unaugmented root convention also allows one-time
kernelization to control the polynomial dependence on the input
size.

The main contribution is a constructive way to use structure
defined by an unknown optimum. The critical-edge representation
connects the optimum gap to locations in the input tree, making
it possible to balance the unresolved disagreement. Target
preservation and marker reconstruction then allow the resulting
sides to be solved independently without losing optimality.
A shared deletion count connects the cost of enforcing separation
with the work left for the side solver. Together, these properties
give an explicit transformation of exponential bases:
a compatible tree--forest solver with base $B>\varphi$ yields
an algorithm with base $\sqrt{\varphi B}$. Since the resulting
algorithm provides the same input and output guarantees, this
improvement can be applied repeatedly. The framework therefore
provides a reusable method for improving MAF algorithms while
keeping the side solver separate from the decomposition and
reconstruction procedures.

An important next question is what determines the limiting base
$\varphi$. In the present analysis, it comes from the weighted
cost of binary separation. Understanding whether this cost is
inherent to the separation requirements, or can be reduced by
a different procedure, would clarify the scope for further
exponential improvements. More broadly, the trade-off between
separation cost and side-gap balance deserves a systematic
analysis. Such an analysis could also identify decompositions
that achieve a given exponential bound with fewer composition
levels and a smaller polynomial factor.

Another direction is to identify sufficient conditions for
similar solver improvements in other agreement-forest models.
The central questions are whether the optimum gap admits a
small structural witness that can be balanced, and whether
the resulting boundaries can be recorded and reconstructed
with controlled changes in forest order. Establishing these
properties for unrooted, multifurcating, or multiple-tree inputs
would determine how far the composition principle developed
here can be extended.

\section*{Author contributions}
Feng Shi: Conceptualization, methodology, software.
Jianxin Wang: Data curation, writing---original draft.

\clearpage
\FloatBarrier
\appendix

\section{Encoding the Distinguished Root Label}
\label{app:root-encoding}

The rSPR characterization in Section~\ref{sec:maf-rspr} places
$\rho$ at the root of each augmented tree. The following encoding
lets us apply our algorithm, which takes trees with all labels
on leaves.

\begin{lemma}
\label{lem:root-leaf-encoding}
Let $T_1,T_2$ be $X$-trees and let $\rho\notin X$. For each
$i\in\{1,2\}$, let $T_i^\rho$ be obtained by adding a root
labeled $\rho$ whose only child is the original root of $T_i$.
Form the $(X\cup\{\rho\})$-tree $\overline T_i$ by moving
the label $\rho$ from the added root to a new leaf attached
to that root.
A partition of $X\cup\{\rho\}$ is the component label partition
of an \af for $(T_1^\rho,T_2^\rho)$ if and only if it is the
component label partition of an \af for
$(\overline T_1,\overline T_2)$. In particular,
$m(T_1^\rho,T_2^\rho)=m(\overline T_1,\overline T_2)$.
An \af for either pair can be converted in polynomial time to
one for the other pair with the same component label sets.
\end{lemma}

\begin{proof}
Fix a partition of $X\cup\{\rho\}$ and let $B$ be the block
containing $\rho$. For each $i$, every other block has the same
minimal connected subgraph in $T_i^\rho$ and $\overline T_i$,
entirely within $T_i$. Write $Y=B\setminus\{\rho\}$. If
$Y=\varnothing$, the subgraph for $B$ is a single vertex
outside $T_i$ in each tree.
Otherwise, in both trees its part inside $T_i$ is the union
of the paths from the original root to the labels in $Y$.
This includes the path from the original root to the least
common ancestor of $Y$. Since only $B$ can use vertices outside
$T_i$, the minimal subgraphs for the blocks are pairwise
vertex-disjoint in $T_i^\rho$ exactly when they are in
$\overline T_i$.

Forced contraction gives the same rooted restriction for
each block contained in $X$. If $Y=\varnothing$, the restriction
for $B$ is the singleton $\{\rho\}$ in both trees. If
$Y\ne\varnothing$, forced contraction suppresses every vertex
strictly between the added root and the least common ancestor
of $Y$. The restriction for $B$ in $T_i^\rho$ then has a root
labeled $\rho$ with the root of $T_i[Y]$ as its only child.
In $\overline T_i$, the restriction for $B$ has an unlabeled
root with two children, the leaf labeled $\rho$ and the root
of $T_i[Y]$. For nonempty $Y$, both encodings therefore give
matching restrictions for $B$ exactly when $T_1[Y]$ and
$T_2[Y]$ are isomorphic. This proves the equivalence of the
\af conditions.

To convert an \af, locate the component containing $\rho$.
If it has other labels, move $\rho$ from its root to a new
child leaf, or reverse the operation by deleting the leaf
labeled $\rho$ and labeling its parent $\rho$. A singleton
component $\{\rho\}$ needs no change. This conversion takes
polynomial time and preserves the order.
\end{proof}

\section{Correctness of the Reduction Rules without Root Augmentation}
\label{app:kelk-compatibility}

We prove Lemma~\ref{lem:kelk-unaugmented-compatibility}. For each
rule, we establish how the \maf order changes and how to
recover an \af for the original pair from one for the reduced pair.
Throughout, $(T_1,T_2)$ is the original pair and $(S_1,S_2)$ is
the pair obtained by applying the rule under consideration.

\subsection{Subtree reduction}

Let $U$ be the label set of the common pendant subtree, and let
$x\in U$ be the representative label. Restricting an \af for
$(T_1,T_2)$ to $(X\setminus U)\cup\{x\}$ gives an \af for
$(S_1,S_2)$ of no larger order. Hence
$m(S_1,S_2)\leq m(T_1,T_2)$.

Conversely, take an \af for $(S_1,S_2)$ and expand the leaf $x$
in its component into the common pendant subtree. In both inputs,
the expansion replaces the same labeled leaf in the common rooted
component by the same rooted tree. All other components have
labels outside $U$, so their minimal representations in the
original inputs avoid the pendant subtree. The expansion therefore
gives an \af for $(T_1,T_2)$ of the same order. In particular,
$m(T_1,T_2)\leq m(S_1,S_2)$, and expanding a reduced \maf gives
a \maf for $(T_1,T_2)$.

\subsection{Common-chain reduction}

Let $(S_1,S_2)$ be obtained from $(T_1,T_2)$ by shortening
a maximal common chain $(x_1,\ldots,x_\ell)$ with $\ell\geq4$
to $(x_1,x_2,x_3)$ as in Section~\ref{sec:root-kernel-discussion}.
To restore the deleted labels, we first put the three retained
labels in a single component of the returned forest.

\begin{lemma}
\label{lem:common-chain-normalization}
Given any \af $H$ for $(S_1,S_2)$, we can construct in
polynomial time an \af $H'$ for $(S_1,S_2)$ such that
$\ord(H')\leq\ord(H)$ and $x_1,x_2,x_3$ lie in the same
component of $H'$.
\end{lemma}

\begin{proof}
For two trees, the chain rule of~\cite{kelk2026multiplekernel}
allows any three chain labels to be retained, so it applies
when we keep $x_1,x_2,x_3$. The block splits and merges from
Lemma~2 of that paper, as adapted to rooted trees in Lemma~5,
give an \af $H'$ for $(S_1,S_2)$ with
$\ord(H')\leq\ord(H)$ and all three retained labels in one
component. We can perform these operations in polynomial time
by inspecting the component representations at the ends of
the retained chain and updating the affected blocks.
\end{proof}

Restricting an \af for $(T_1,T_2)$ to the retained labels gives
an \af for $(S_1,S_2)$ of no larger order. Thus
$m(S_1,S_2)\leq m(T_1,T_2)$.

For the converse, apply Lemma~\ref{lem:common-chain-normalization}
to an \af $H$ for $(S_1,S_2)$, and let $C$ be the component
of $H'$ containing $x_1,x_2,x_3$.
View the representations of $H'$ in the original inputs by
expanding suppressed paths, and add $x_4,\ldots,x_\ell$ to
the label set of $C$. In each original tree, these labels
attach along the chain immediately above the parent of $x_3$.
In the reduced component, this parent either has an incoming
edge or is the root of $C$. The restored labels therefore
attach in the same order along that edge, or above the root,
in both inputs. Any path between labels of another component
that uses this restored segment would also pass through the
parent of $x_3$, which already belongs to the representation
of $C$. Thus the restored component has the same rooted
topology in both inputs and remains disjoint from the other
components in each input.

Reinsertion takes polynomial time and leaves the number of
components unchanged. The resulting \af for $(T_1,T_2)$
therefore has order at most $\ord(H)$ and contains the entire
chain in one component. Taking $H$ to be a \maf yields
$m(T_1,T_2)\leq m(S_1,S_2)$, so the two minimum orders are equal.

\subsection{\texorpdfstring{The $3$-$2$-chain reduction}{The 3-2-chain reduction}}

We adapt the component exchanges of Kelk et
al.~\cite[Lemma~5.1]{kelk2023cyclic}. Write $a=x_i$,
$b=x_j$, and $c=x_3$, with $b$ the deleted label.
Assume that $T_1$ has the pendant subtree $((a,b),c)$
and that $T_2$ has the cherry $(a,c)$.

Take an \af for $(S_1,S_2)$ and keep its component label sets.
In each original tree $T_i$, the representation of each component
is obtained by replacing every edge of its representation in
$S_i$ with the corresponding path in $T_i$. This preserves its rooted
topology after forced contraction. The representations remain
vertex-disjoint because the paths corresponding to distinct
edges have disjoint interiors. They all avoid the leaf $b$,
so adding $\{b\}$ as a singleton gives an \af for $(T_1,T_2)$
with one additional component. Thus
$m(T_1,T_2)\leq m(S_1,S_2)+1$.

For the reverse inequality, let $G$ be a \maf for $(T_1,T_2)$
and work with its component label sets. We show that there is a
\maf with $\{b\}$ as a block. If $b$ is already a singleton,
we are done. Otherwise let $B$ be its block, with $|B|\geq2$.
The two trees induce different rooted triples on $\{a,b,c\}$,
so $B$ contains at most one of $a$ and $c$.

First suppose that $B$ contains exactly one of $a,c$, denoted by
$u$, and let $v$ be the other. Then $v$ is a singleton.
If $u=c$, the path from $b$ to $c$ in $T_1$ passes through the
parent of $a$. If $u=a$, the path from $b$ to $a$ in $T_2$
passes through the parent of $c$. In each case, any representation
joining $v$ to another label would meet the representation of $B$.

Replace the blocks $B$ and $\{v\}$ by
$B'=(B\setminus\{b\})\cup\{v\}$ and $\{b\}$.
In $T_1[B]$, $b$ and $u$ form a cherry because the triple
is pendant and $v$ is absent from $B$. They also form a cherry
in $T_2[B]$ because $B$ is an agreement block. In $T_1[B']$,
the cherry $(a,c)$ occupies the same position relative to the
labels outside the triple. In $T_2[B']$, the original cherry
$(a,c)$ takes the place of $(b,u)$ in $T_2[B]$. To see this,
first restrict to $B\setminus\{b\}$, which reduces $(b,u)$
to the leaf $u$, and then restore the sibling $v$ of $u$ in
$T_2$. Thus $T_1[B']$ and $T_2[B']$ both have the topology
of the old component with its cherry relabeled.

In $T_1$, the new representation can use additional vertices
only inside the pendant triple. All unchanged blocks have their
labels outside that subtree, so their representations avoid it.
In $T_2$, the parent of $v$ is already on the path from $u$ to
$b$. The new representation is contained in the old one together
with the edge to $v$ and the leaf $v$. Hence the replacement
preserves vertex-disjointness in both trees.

Now suppose that $B$ contains neither $a$ nor $c$. The path
in $T_1$ from $b$ to another label of $B$ uses both internal
vertices of the pendant triple. A component joining either
$a$ or $c$ to another label would meet this path, so both are
singletons. Replace $B,\{a\},\{c\}$ by
$B\setminus\{b\},\{a,c\},\{b\}$. The first block is nonempty
and retains a common rooted restriction. Its representation
avoids the pendant triple in $T_1$ and the cherry $(a,c)$ in
$T_2$, as do the representations of the unchanged blocks.
Thus the new block $\{a,c\}$ has a common two-leaf topology
and a representation disjoint from all other blocks.

Both exchanges preserve the number of blocks and make $b$ a
singleton. The argument uses only vertices within the pendant
triple or cherry and the old component representations, so it
also covers the case in which the pendant triple is all of $T_1$.
Removing the singleton $\{b\}$ and restricting the inputs gives
$m(S_1,S_2)\leq m(T_1,T_2)-1$. Combining the two inequalities gives
$m(S_1,S_2)=m(T_1,T_2)-1$.

\clearpage
\bibliographystyle{plainnat}
\bibliography{mybibfile}

\end{document}